\documentclass[aps,prl,twocolumn,superscriptaddress]{revtex4}
\usepackage{graphicx}
\usepackage{times}
\usepackage{amsmath}
\usepackage{textcomp}

\def\etal{\textit{et al.\xspace}} 
\usepackage{textcomp}
\usepackage{color} 

\newcommand{\subs}[1]{\ensuremath{{}_{\textnormal{#1}}}}

\begin{document}

\title{Magnetotransport and induced superconductivity in Bi based three-dimensional topological insulators }
\author{M. Veldhorst}
\thanks{corresponding author: m.veldhorst@unsw.edu.au}
\altaffiliation[Now at ]{University of New South Wales, Australia} 
\affiliation{Faculty of Science and Technology and MESA+ Institute for Nanotechnology, University of Twente, 7500 AE Enschede, The Netherlands}
\author{M. Snelder}
\affiliation{Faculty of Science and Technology and MESA+ Institute for Nanotechnology, University of Twente, 7500 AE Enschede, The Netherlands}
\author{M. Hoek}
\affiliation{Faculty of Science and Technology and MESA+ Institute for Nanotechnology, University of Twente, 7500 AE Enschede, The Netherlands}
\author{C.G. Molenaar}
\affiliation{Faculty of Science and Technology and MESA+ Institute for Nanotechnology, University of Twente, 7500 AE Enschede, The Netherlands}
\author{D.P. Leusink}
\affiliation{Faculty of Science and Technology and MESA+ Institute for Nanotechnology, University of Twente, 7500 AE Enschede, The Netherlands}
\author{A.A. Golubov}
\author{H. Hilgenkamp}
\affiliation{Faculty of Science and Technology and MESA+ Institute for Nanotechnology, University of Twente, 7500 AE Enschede, The Netherlands}
\author{A. Brinkman}
\affiliation{Faculty of Science and Technology and MESA+ Institute for Nanotechnology, University of Twente, 7500 AE Enschede, The Netherlands}
\date{\today}

\begin{abstract}
The surface of a 3D topological insulator is conducting and the topologically nontrivial nature of the surface states is observed in experiments. It is the aim of this paper to review and analyze experimental observations with respect to the magnetotransport in Bi-based 3D topological insulators, as well as the superconducting transport properties of hybrid structures consisting of superconductors and these topological insulators. The helical spin-momentum coupling of the surface state electrons becomes visible in quantum corrections to the conductivity and magnetoresistance oscillations. An analysis will be provided of the reported magnetoresistance, also in the presence of bulk conductivity shunts. Special attention is given to the large and linear magnetoresistance. Superconductivity can be induced in topological superconductors by means of the proximity effect. The induced supercurrents, Josephson effects and current-phase relations will be reviewed. These materials hold great potential in the field of spintronics and the route towards Majorana devices.
\end{abstract}

\maketitle

\section{Introduction}
\begin{figure}
\centering
\includegraphics[width=0.3\textwidth]{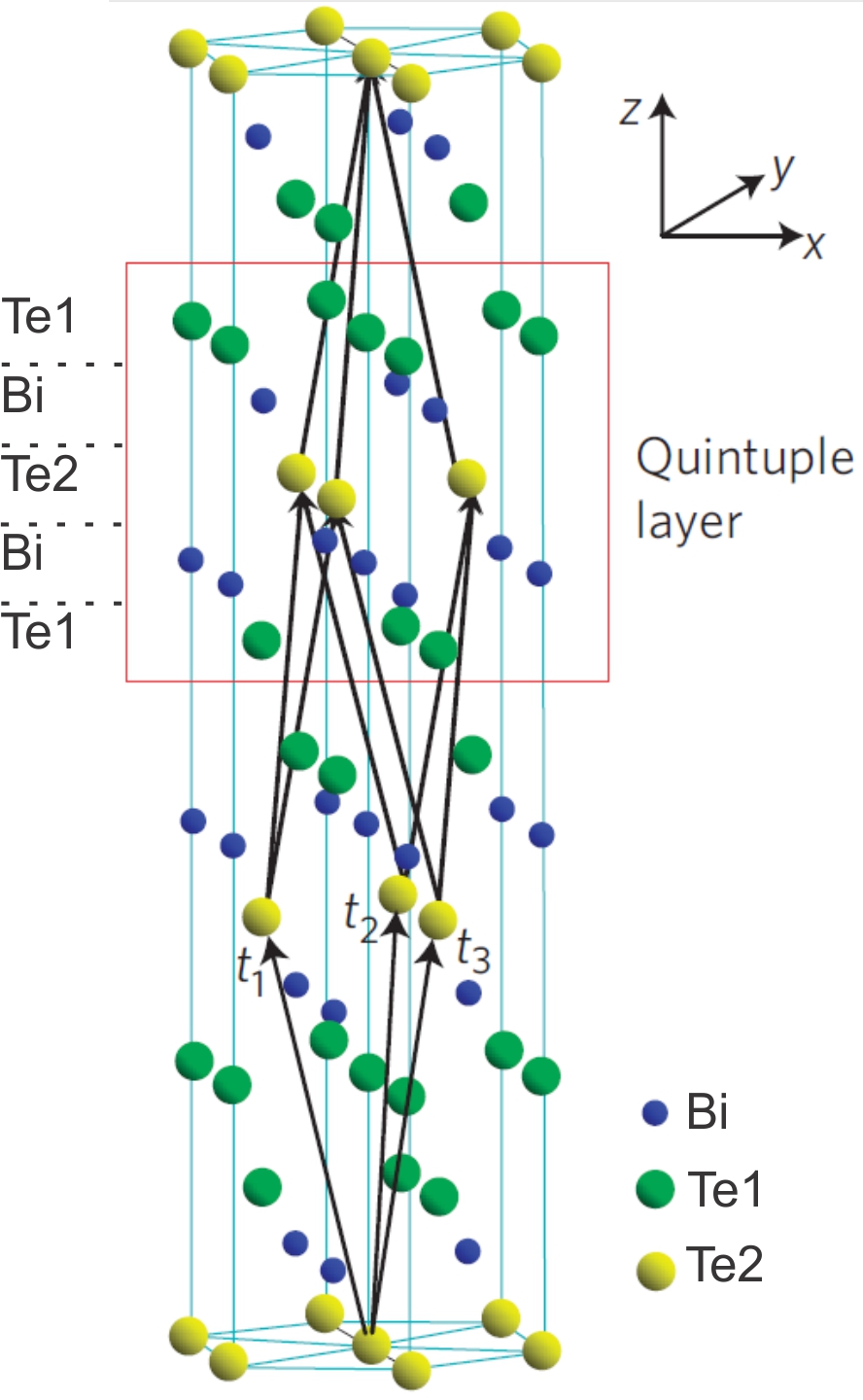}\vspace{-5pt}
\caption{The crystal structure of Bi$_{2}$Te$_{3}$ as a representative example of Bi-based 3D topological insulators. From \cite{Zhang2009}.} 
\label{fig:x}
\end{figure}

A three-dimensional (3D) topological insulator (TI) is a semiconductor with two-dimensional (2D) surface states that have an energy dispersion across the entire bulk bandgap. These surface states arise from band inversion, due to strong spin-orbit interactions. The band inversion makes the surface states topologically nontrivial, meaning that these states are protected (cannot be removed) and that the states have a helical Dirac-type dispersion, where spin is tightly coupled to momentum. Several compounds have been theoretically predicted to be 3D TIs \cite{Fu2007a}. Strong experimental evidence for the presence of topological surface states exists for the class of Bi-based materials: the alloy Bi$_{1-x}$Sb$_{x}$ \cite{Fu2007b}, and the compounds Bi$_2$Se$_3$ and Bi$_2$Te$_3$ \cite{Zhang2009}, see Fig. \ref{fig:x} for the crystal structure. The class of Bi-based topological insulator has been expanded with the fabrication of TlBiSe$_2$ \cite{Sato2010,Kuroda2010,Chen2010b}, TlBiTe$_2$ \cite{Chen2010b}, BiSbPbTe \cite{Souma2012}, and PbBiTe$_4$ \cite{Kuroda2012}. The bulk of these compounds is not necessarily insulating, due to the presence of defects and impurities. This bulk conductivity might shunt and mask topological transport properties. A successful trend in the research on topological insulators has been to reduce the number of defects and impurities by atom substitutions such as Bi$_{2}$Te$_{2}$Se\cite{Xu2010,Ren2010,Xiong2011} and Bi$_{2-x}$Sb$_x$Te$_{3-y}$Se$_y$  \cite{Ren2011}. which have higher bulk resistivities. Simultaneously, progress has been made in disentangling topological surface state transport properties from bulk transport contributions.
Here, we review and analyze the experimentally observed magnetotransport properties of the Bi based 3D TIs as well as the induced superconductivity in these compounds. Studying transport properties in an externally applied large magnetic field is a powerful probe of the topological nature of the surface states. A Berry phase shift appears in magnetoresistance oscillations and in quantum corrections to the conductivity, leading for example to weak anti-localization effects. We will review the ongoing debate on the manifestation of this phase shift and show how the phase shift depends on the actual parameters of the sample. We will discuss the important role of bulk conductivity contributions. Intriguingly, two-dimensional transport features not always originate from the surface in Bi-based topological insulators. This finding is also important for understanding the presence of a large and linear magnetoresistance in most Bi-based 3D TIs.

The topologically non-trivial surface state is an attractive medium to study unconventional superconductivity. When superconductivity exists in a system with a Dirac-type dispersion and helical spin-momentum locking, exotic features could be realized such as $p$-wave order parameter symmetry \cite{Kitaev2001} and Majorana type bound states \cite{Fu2008}. Josephson hybrid structures have been designed theoretically already to mimic non-abelian particle statistics \cite{Fu2008}. Parallel to the trend of doping and alloying TIs to render them superconducting \cite{Sasaki2011,Heumen2012} runs the successful approach of inducing superconductivity from a standard superconductor into a TI by means of the proximity effect. Here, we will review the current status of the experimental efforts in the latter direction. Evidence for supercurrents through topological surface states will be analyzed, also in the presence of bulk conducting channels. The prospects for future SQUID based experiments to detect topological effects in the superconductivity will be discussed.

\section{Weak antilocalization}
\begin{figure*}
\centering
\includegraphics[width=0.95\textwidth]{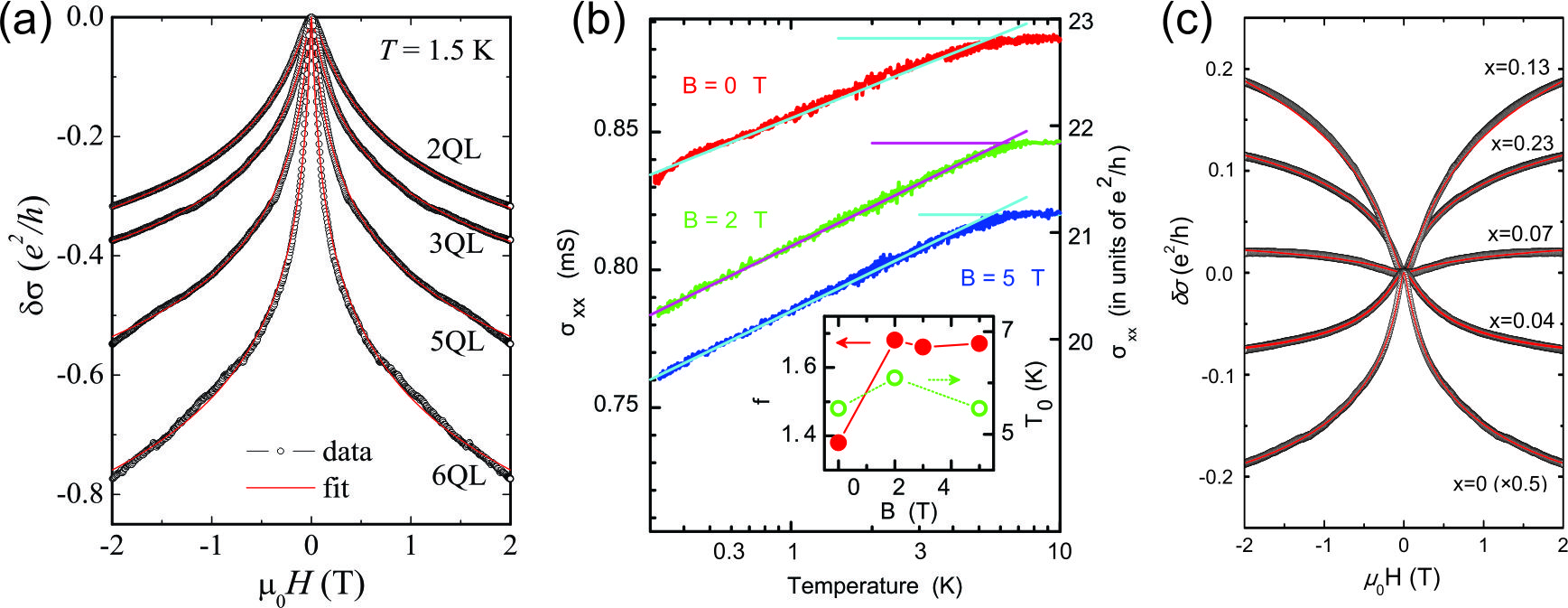}\vspace{-5pt}
\caption{Weak antilocalization in topological insulator thin films and crystals. (a) thickness dependence of WAL for thin Bi$\subs{2}$Se$\subs{3}$ films by Liu \textit{et al.}\cite{Liu2011}, the red line is a fit with the HLN equation\cite{Hikami1980}, Eq. (\ref{eq:HLN}). (b) Temperature dependence in magnetic field of an 80 nm Cu$_x$Bi$_{2-x}$Se$_3$ film ($x$ = 0.02 - 0.03) \cite{Takagaki2012}. The logarithmic temperature dependence is largely field independent. The inset shows $f$ and $T_0$ determined from a fit to Eq. (\ref{eq:EEI}), where $f=\left(1-\frac{3}{4}F\right)$ and $T_0=T_{ee}$. The WAL contribution is suppressed above 2 T.  (c) WAL to WL crossover observed in Bi$_{2-x}$Cr$_x$Se$_3$ films as a function of Cr doping \cite{Liu2012}. Red lines are fits using the HLN equation.} 
\label{fig:WAL}
\end{figure*}

Weak antilocalization (WAL) is commonly observed in magnetoresistance measurements on thin films or flakes of Bi-based 3D TIs, see Fig. \ref{fig:WAL}(a). WAL is observed for a wide range of thicknesses. Kim \textit{et al.}\cite{Kim2011} report magnetotransport on Bi$\subs{2}$Te$\subs{3}$ films from 3 nm to 170 $\hbox{\textmu}m$ thick, where they can still see the WAL effect. When caused by topological surface state transport, WAL may more properly be named topological delocalization\cite{Liu2011}, because the antilocalization is caused by the $\pi$ -Berry phase (see section 3) picked up along two time-reversed self-intersecting paths. Effectively, in TI surface states all the WL orbits contribute as WAL orbits. WL will not appear in TIs, as long as time reversal symmetry is not broken.

The Hikami-Larkin-Nagaoka (HLN) equation\cite{Hikami1980} describes the quantum correction to 2D conductivity due to (anti-)localization effects as
\begin{equation}
\begin{array}{l}
\Delta \sigma_{xx}(B)=  \sigma_{xx}(B)-\sigma_{xx}(0)\\
=   \alpha\cdot\frac{e^2}{2\pi^2\hbar}\left[\Psi\left(\frac{1}{2}+\frac{B_{\phi}}{B}\right)-\ln\left(\frac{B_{\phi}}{B}\right)\right],
\end{array}
\label{eq:HLN}
\end{equation}
where $\Psi$ is the digamma function, $B_{\phi}=\frac{\hbar}{4eL_{\phi}^2}$, and $L_{\phi}$ is the dephasing length. Strong spin-orbit interactions (or topological effects) give WAL with $\alpha = -0.5$ while weak spin-orbit scattering typically gives weak localization (WL) effects with $\alpha = 1$. A logarithmic temperature dependence is expected, since $L_{\phi}$ generally scales inversely with temperature. For WL the conductivity decreases with decreasing temperature and for WAL the conductivity increases with decreasing temperature. An applied magnetic field will reduce the effect of the quantum corrections, due to a phase being picked up in the magnetic field, destroying the interference effects. Equation (\ref{eq:HLN}) is an approximation of an expression that takes all the scattering lengths explicitly into account \cite{Hikami1980}. 

For topological insulator devices, the value of $\alpha$ in Eq. (\ref{eq:HLN}) is usually around $-0.5$, but it is found to depend on sample thickness, the presence of bulk conductivity, gating and the substrate, as will be elaborated below. The trivial value of $\alpha =0$ is only found for higher temperatures or for very thin samples in a parallel applied field \cite{He2011}. Depending on the number of conduction channels, $\alpha$ can even become $-1$ when both the top and the bottom surface of a crystal or film contribute independently \cite{Zhang2011a,Steinberg2011,Checkelsky2011}. 
The 2D nature of the WAL is confirmed by angle dependent measurements. For a 2D system, the WAL effect only depends on the perpendicular component of the field \cite{He2011,Cha2012a}. For higher fields extra spin effects may arise due to the Zeeman splitting caused by the parallel field \cite{Iordanskii1994,Meijer2005}.

\subsection{Bulk \textit{vs} surface}
Even when the 2D nature of the WAL is confirmed by angle dependent measurements, the WAL can still be caused by bulk effects. Quite a number of reports \cite{Kim2011,Liu2011,He2011,Zhang2011a,Steinberg2011,Wang2011,Takagaki2012,Zhang2012,Chen2010,Chen2011,Taskin2012,Gao2012} deduce a dephasing length from the WAL fit using the HLN-equation that is larger than the film thickness, making the bulk effectively more 2D. Also, it is an intriguing question why in most cases the observed WAL can only be explained by just one of the two surfaces (i.e. $\alpha=-0.5$). One option would be that the whole topological insulator can be considered as one 2D system with spin-orbit coupling, as a mix of bulk channels and surface states \cite{He2011,Zhang2011a,Steinberg2011,Chen2011,Taskin2012}. The question then is how these contributions can be disentangled. An answer might be found in the hints of bulk contribution found in some works, such as a broader WAL from the bulk \cite{Cha2012a}, WL channels \cite{Zhang2012b,Lu2011,Garate2012} or by specifically trying to influence the surface states, for example by gating \cite{Zhang2011a,Steinberg2011,Checkelsky2011,Gao2012} or by introducing magnetic particles \cite{He2011}. A theoretical description of the bulk-surface coupling is given by Refs. \cite{Lu2011} and \cite{Garate2012}.

He \textit{et al.} \cite{He2011} show that the WAL in Bi$_2$Te$_3$ thin films has contributions from both the surface and the bulk, by analyzing angle dependent measurements. The surface contribution to the WAL can be suppressed by depositing magnetic impurities (Fe) on the surface. For very thin films they observe a crossover to the regime of a $B^2$ field dependence. The question does arise what happens to the bottom interface; the film is smaller than the phase coherence length and the bottom interface should not be affected by the magnetic impurities on the top interface. The authors note that the contributions from the bottom surface seem negligible and that this may be caused by defects present at the TI-substrate interface. Checkelsky \textit{et al.} \cite{Checkelsky2011} find contributions from two interfaces using the HLN equation to fit the WAL measured on exfoliated flakes of Bi$_2$Se$_3$, while still having a phase coherence length larger than the thickness of the flakes. This gives merit to the claim that for grown thin films, the bottom interface has too much disorder to show a WAL effect \cite{He2011}.

Steinberg \textit{et al.} \cite{Steinberg2011} observe a tunability of the number of channels leading to WAL in top gate tuned 20 nm Bi$_2$Se$_3$ thin films, both by temperature and applied electric field. The authors argue that the different channels can be viewed as independent as long as carriers in one channel lose coherence before scattering to the other channel, i.e. $\tau_{SB}>\tau_{\phi}$, where $\tau_{SB}$ is the surface-to-bulk scattering time. This can be the case when a depletion layer is formed between the surface and the bulk by applying a gate voltage \cite{Zhang2011a,Steinberg2011,Checkelsky2011,Chen2011,Gao2012}. Steinberg \textit{et al.} \cite{Steinberg2011} observe that the HLN fit also works in the crossover regime between $\alpha=-1$ and $-0.5$, indicating that, while there is no theoretical model, $\alpha$ can be used as a phenomenological measure for the channel separation (i.e. the ratio $\tau_{SB}/\tau_{\phi}$). As the temperature is increased the channel separation is seen to increase, which is explained by the strong temperature dependence of $\tau_{\phi}$ versus the relatively temperature independent $\tau_{SB}$, which is suggested to be governed by impurity scattering \cite{Steinberg2010}. A similar empirical model is suggested by Kim \textit{et al.} \cite{Kim2011}. In their measurements on up to 100 unit cells of Bi$_2$Se$_3$, the mobility depends linearly on the film thickness, indicating that $L_{\phi}$ should also depend linearly on film thickness if the WAL would be a purely bulk effect. However, in the thin film limit the scaling of $L_{\phi}$ deviates from a linear scaling, indicating it cannot be explained by only a bulk contribution. The authors note that the scaling of $L_{\phi}$ with thickness $t$ can be used as a figure of merit for TIs. For $L_{\phi} \propto t^{s}$, $s$ should be zero for a perfect TI with insulating bulk, one for a trivial strong spin-orbit coupling material or somewhere in between for a TI with both surface and bulk contribution.

\subsection{Localization to antilocalization crossover}
A combination of WL and WAL is observed at low temperatures in ferrocene doped Bi$_2$Se$_3$ nanoribbons \cite{Cha2012a}. This can be explained by the opening of a band gap in the surface states induced by the magnetic impurities \cite{Lu2011a} or by a canceling of the bulk spin-orbit coupling by the magnetic impurities \cite{Cha2012a}. A more detailed study with controlled magnetic doping is needed to disentangle these two effects. A coexistence of WL and WAL for nonmagnetic dopants has not been observed \cite{He2011,Cha2012a,Cha2012}, which is an indication that (a significant portion of) the WAL is caused by the topologically protected surface states.

Liu \textit{et al.} \cite{Liu2012} observe a crossover as a function of Cr doping in Bi$_2$Se$_3$ films of 3 unit cell quintuple layers (QL) grown by molecular beam epitaxy on sapphire, see Fig. \ref{fig:WAL}(c). While this thickness is already in the regime where scattering between the two surfaces opens up a gap in the surface states \cite{Zhang2010}, the authors note that the observed effects are still nominally the same as in 6 QL films, but without a quadratic bulk contribution. The crossover looks similar to what has been predicted theoretically\cite{Lu2011a}, driven by the opening of a gap due to the magnetic impurities \cite{Liu2009,Chen2010a}. However, the authors note that from their ARPES measurements, it looks like the surface states completely disappear for large magnetic doping, leaving the system in a dilute magnetic semiconductor state. They find $\alpha = 0.14$ for the HLN fit for the highly doped films, which is indeed closer to the regime of strong magnetic scattering than the WL regime \cite{Hikami1980}. The authors observe a further crossover as a function of temperature, where for the highly doped samples there is a coexistence of WL and WAL, each with a different critical field scale \cite{Liu2012}. This coexistence could be another explanation for the low value of $\alpha$; in the crossover regime the limiting cases used to obtain Eq. (\ref{eq:HLN}) do not hold and the full expression should be used \cite{Hikami1980,Bergmann1984}.

\subsection{Electron-electron interactions}
Electron-elec\-tron interactions (EEI) can lower the conductivity via the Aronov-Altshuler effect \cite{Altshuler1979,Lee1985}. The Coulomb interaction between different electrons decreases the effective density of states near the Fermi energy, which results in a correction to the conductivity that is logarithmic in temperature. In magnetic fields the conductivity will further decrease as a result of Zeeman splitting\cite{Lee1985}.

The correction to the conductivity by EEI \cite{Lee1985} is given by 
\begin{equation}
\Delta\sigma_{ee}(T)=-\frac{e^2}{\pi h}\left(1-\frac{3}{4}F\right)\ln\left(\frac{T}{T_{ee}}\right),
\label{eq:EEI}
\end{equation}
where $F$ is the Coulomb screening factor and $T_{ee}$ is the characteristic temperature for EEI effects.

EEI is used by some authors \cite{Liu2011,Takagaki2012,Checkelsky2009} to explain a decrease in conductance that depends logarithmically on temperature, where an increase is expected for just WAL. This decrease is not always observed: in some cases there is only a metallic temperature dependence \cite{Zhang2011a,Taskin2012}. EEI have already been used to explain localization effects in other strong spin-orbit materials \cite{Beutler1988,Lin1987}. The EEI contribution to conductivity is extracted from the conductance in magnetic fields. 

Checkelsky $et$ $al.$ \cite{Checkelsky2009} deduce both orbital and spin contributions to the magnetotransport from the angle dependence of universal conductance fluctuations. The low field effect is fit by a combination of WAL \cite{Hikami1980} (the orbital part) and EEI \cite{Lee1985} (the spin part). They find a logarithmic field dependence, but the fitting parameters are orders of magnitude larger than expected for a 2D system. The authors reason that extra 2D bulk states with similar spin-orbit properties as the surface states may be able to explain this discrepancy \cite{Analytis2010a}.

Takagaki $et$ $al.$ \cite{Takagaki2012} determine the prefactor and $T_{ee}$ in Eq. (\ref{eq:EEI}) by looking at the temperature dependence in magnetic fields, to suppress the contribution of WAL, see Fig. \ref{fig:WAL}(b). The value they find for the screening factor is too large to only come from the surface, which suggests that the bulk states also have to be taken into account. 

Liu $et$ $al.$ \cite{Liu2011} observe a combination of WAL and EEI in ultrathin Bi$_2$Se$_3$ films of one to six QL grown on sapphire. The logarithmic temperature dependence is determined to be caused by EEI by studying the MR in parallel fields. EEI is largely field independent, but for large Zeeman splitting it will be suppressed with a logarithmic field dependence. The g-factor is not very large in TIs \cite{Xiong2012a}, but Liu $et$ $al.$ indeed observe a logarithmic field dependence for fields above 4 T. They argue that the EEI becomes larger for thinner films due to increased disorder and reduced dimensionality, leading to poorer screening and therefore stronger Coulomb interaction \cite{Liu2011}. 

\section{Shubnikov-de-Haas oscillations and Berry phase}
The mobility $\mu$ of the Bi-based topological insulators has frequently been reported to be high enough to fulfill the condition $\mu B \gg 1$, where Shubnikov-de-Haas oscillations can be observed in the resistance at high magnetic fields \cite{Ren2010,Taskin2012,Xiong2012a,Qu2010,Taskin2010,Analytis2010,Sacepe2011,Brune2011,Xiu2011,Taskin2011b,Xiong2012,Veldhorst2012a,Cao2012,Ren2011r,Ren2012}. The oscillations in the longitudinal surface resistance $\rho_{xx}^S$ are given \cite{Lifshitz1956} by 
\begin{equation}
\Delta \rho_{xx}^S \propto \textrm{cos} \left( 2 \pi \frac{F}{B} + \pi + \phi_B \right),
\label{SdH}
\end{equation}
where $\phi_B$ is a Berry phase, as will be discussed below, and the frequency $F$ is determined from the extremal cross-section of the Fermi surface, $2\pi F=\hbar S(E_F) /e$. 

If the electron orbit is confined to a 2D plane as for a two-dimensional electron gas, then $B$ is the perpendicular magnetic field component. However, a cosine magnetic field angle dependence of the oscillations alone is not enough to conclude that the topological surface states are providing the oscillations. For example, it has been shown for Bi$_2$Se$_3$ that the bulk may act as the sum of many parallel 2D electron systems \cite{Cao2012}. And a surface state might also be a non-topological trivial band that crosses the Fermi energy due to band bending at the surface, as observed for Bi$_{1.5}$Sb$_{0.5}$Te$_{1.7}$Se$_{1.3}$ \cite{Taskin2011b}. 

Rather, the topological nature can be concluded from the phase of the Shubnikov-de-Haas oscillations. From graphene it is known how a Dirac cone gives rise to a half filled Landau level at zero energy (here defined as the energy of the Dirac point) and how this Landau level gives an additional Berry phase shift of $\pi$ to the quantum Hall effect or Shubnikov-de-Haas oscillations \cite{Novoselov2005,Zhang2005}. In general, the Berry phase can be obtained by integrating the Berry connection $\boldsymbol{\Omega}$ over an electron orbit
\begin{equation}
\phi_B = \oint \boldsymbol{\Omega} d \textbf{k},
\label{Berry}
\end{equation}
where the Berry connection follows from a specific band structure as
\begin{equation}
\boldsymbol{\Omega}= i\int u_{\textbf{k}}^{\ast} (\textbf{r}) \boldsymbol{\nabla_k}u_{\textbf{k}} (\textbf{r}) d \textbf{r}, 
\end{equation}
where $u_{\textbf{k}} (\textbf{r})$ are the periodic parts of the Bloch wave functions. It turns out that Eq. (\ref{Berry}) gives $\phi_B = \pi$ at the surface of the Bi-based topological insulators irrespective of the exact shape of the Dirac cone \cite{Mikitik2012}, i.e. deviations of the linear dispersion in a Dirac cone do not lead to different values for the Berry phase, in contrast to previous expectations \cite{Taskin2011}. The crucial  ingredient to have a $\pi$ Berry phase is that the order of conduction band and valence band are inverted due to strong spin-orbit coupling. 

For the trivial $\phi_B=0$ case, Eq. (\ref{SdH}) shows that the magnetoresistance oscillations have a minimum in the limit $1/B\rightarrow 0$. A plot of the $1/B$ positions as function of the $n^{\textrm{th}}$ number of minima then extrapolates to 0. When $\phi_B=\pi$, the $1/B$ positions of the minima extrapolate to $-1/2$. The extrapolation becomes very accurate either when the applied magnetic fields are so large that the lowest Landau levels are resolved, or when an additional gate voltage is used to sweep the Fermi energy \cite{Sacepe2011}. 

Inverting the resistivity tensor gives the longitudinal conductivity as $\sigma_{xx}={\rho_{xx}}/\left({\rho_{xx}^2+\rho_{xy}^2}\right)$. In a one-band free electron model, the oscillation condition $\mu B \gg 1$ coincides with $\rho_{xy} \gg \rho_{xx}$ since $\rho_{xx}=(en\mu)^{-1}$ and $\rho_{xy}=(en)^{-1}B$, where $n$ is the carrier density. When the oscillations in $\rho_{xx}$ are small ($\Delta \rho_{xx} \ll \rho_{xx}$) and when oscillations in $\rho_{xy}$ can be neglected, the fulfilled condition $\rho_{xy} > \rho_{xx}$ automatically ensures that $\Delta \sigma_{xx} \propto \Delta \rho_{xx}$, i.e. the oscillations in conductivity and resistivity are in phase. Both $\sigma_{xx}$ and $\rho_{xx}$ have minima at integer Landau level index $n$. Magnetic field measurements on semiconductors are usually in this regime. 
\begin{figure}[h!]
\centering
\includegraphics[width=0.36\textwidth]{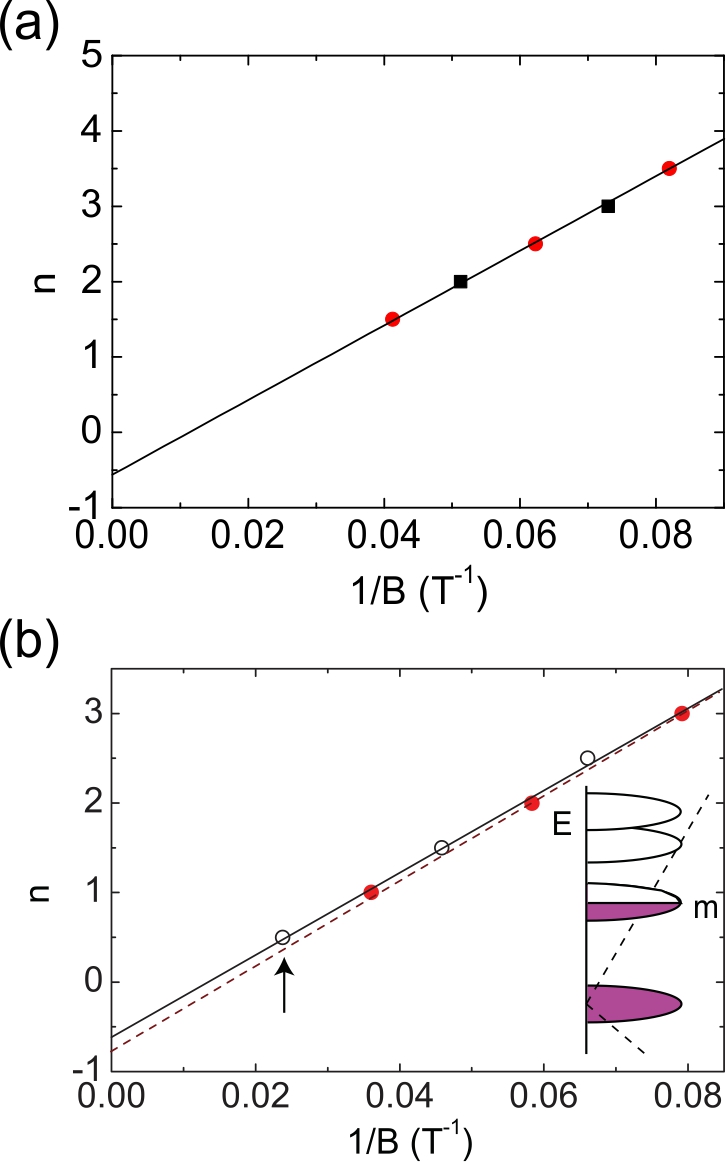}
\caption{Landau level index $n$ as function of $1/B$ for Bi$_{2}$Te$_{3}$ and Bi$_{2}$Te$_{2}$Se. (a) For thin flakes of Bi$_{2}$Te$_{3}$ the Shubnikov de Haas oscillations are shunted by a bulk conductivity with a moderate mobility. In this case, minima positions (black squares) of $\rho_{xx}$ coincide with integer $n$ and extrapolation to $1/B=0$ provides a Berry phase of $\pi$, indicating the topological nature of the surface states. Data is taken from Veldhorst $et$ $al.$ \cite{Veldhorst2012a}. (b) For a large crystal of Bi$_{2}$Te$_{2}$Se with a bulk conductivity shunt with a low bulk mobility, the integer values for $n$ appear for the maxima positions (filled symbols) of $\rho_{xx}$. The inset shows the Landau level filling at 42 T (arrow). From \cite{Xiong2012a}. }
\label{fig:SdH}
\end{figure}

However, in the presence of a bulk shunt with a low mobility, the oscillations in $\rho_{xx}$ and $\sigma_{xx}$ can be out of phase. It was shown for a Bi$_{2}$Te$_{2}$Se crystal showing bulk conductivity with a low mobility that the surface state resistivity has maxima now at integer $n$ \cite{Xiong2012a}. Consequently, the Berry phase needs to be inferred from a plot of the $1/B$ maxima positions at integer $n$, see Fig. \ref{fig:SdH}(b), where the topologically non-trivial $\phi_B=\pi$ was observed. However, this situation \cite{Xiong2012a} is not generic for any strong bulk conductivity shunt. In a parallel summation of surface and bulk conductivity, it can be shown that $\sigma_{xx}$ and $\rho_{xx}$ are in phase when the bulk mobility is not too low, i.e. as long as $\sigma_{xy}^B>\sigma_{xx}^B$.  For example, Cao \textit{et al.} \cite{Cao2012} have shown that the bulk in highly doped Bi$_2$Se$_3$ can act as the sum of many parallel layers with high mobility and a trivial Berry phase with minima at integer $n$. Lower bulk densities will enhance the surface state conduction. In Fig. \ref{fig:SdH}(a) data is shown for a 100 nm thick exfoliated flake of Bi$_{2}$Te$_{3}$. A multiband analysis (assuming two surfaces and one bulk channel) of the Hall data and Shubnikov de Haas oscillations of Ref. \cite{Veldhorst2012a} provides a bulk carrier density of $1.9 \times 10^{19}$ cm$^{-3}$ and a bulk mobility of the order of 0.1 m$^2$V$^{-1}$s$^{-1}$, so that $\sigma_{xy}^B>\sigma_{xx}^B$ is fullfilled above about 10 T. The extrapolation of the minima in $\rho_{xx}$ at integer $n$ consistentently extrapolate to $\phi_B=\pi$ again. For every sample with bulk conductivity a detailed parameter estimate should decide in which regime the oscillations are and whether maxima or minima positions should be considered.

\section{Linear magnetoresistance}
In addition to quantum transport phenomena such as Shubnikov-de Haas oscillations and weak antilocalisation,  a strong background magnetoresistance (MR) was also observed in the 3D TIs Bi$_2$Te$_3$ \cite{Qu2010,Wang2012}, Bi$_2$Se$_3$ \cite{Zhang2011a,Tang2011,He2012,Gao2012}, and  Bi$_2$Te$_2$Se \cite{Assaf2012}. The reported magnetoresistance varies from a few percent to more than 600\% \cite{Wang2012}. In contrast to the more standard small quadratic MR (e.g. due to multiband effects), which saturates for $\mu B>1$, the observed MR is linear with field and does not seem to saturate in high fields. The MR is positive, also for negative field directions, thereby ruling out Hall component admixtures due to misaligned electrodes. 

Linear MR is not a new phenomenon and was already observed about 100 years ago in Bi \cite{Kapitza1928,Abrikosov2000}. The so-called Kapitza linear MR as observed in metals with open Fermi surfaces require large Fermi areas (and therefore large carrier densities) and this does not really apply to any of the Bi-based 3D TIs of current interest. Moreover, more quantitative treatments of open orbits seem to indicate $B^{2/3}$ or $B^{4/3}$ rather than linear behavior. Abrikosov proposed a quantum MR model to account for linear MR \cite{Abrikosov2000,Abrikosov1969,Abrikosov1998} when only one Landau magnetic band is filled. The conditions for the sole occupation of the $n=0$ lowest Landau level (at energy $E_{0}$) is that both the Fermi energy, $E_F$, and $k_BT$ are much smaller than $E_{1}-E_{0}$, where $E_1$ is the energy of the next Landau level. This is called the extreme quantum limit and the resulting linear MR is called quantum linear MR.

Usually, the conditions for the extreme quantum limit are only fulfilled for narrow gap semiconductors or semi-metals with very small Fermi pockets  and low effective masses. Examples are Bi, n-type doped InSb \cite{Hu2008}, and PbS \cite{Eto2010}. For a parabolic dispersion, the Landau levels are given by $E_n=\frac{e\hbar B}{m}\left(n+\frac{1}{2}\right)$. A low effective mass gives a large cyclotron frequency, $\omega_c=\frac{eB}{m}$ and a large spacing between Landau levels. Entering the extreme quantum limit is also facilitated in materials with a linear dispersion. E.g., for $\beta$-Ag$_2$Te a linear MR was observed for fields as low as 10 Oe at 4.5 K \cite{Xu1997}, for which the conditions of the extreme quantum limit do not seem to be fulfilled ($k_BT \ll \hbar \omega_c$ for $B=10$ Oe and $m=0.01 m_e$ gives $T < 0.1$ K). However, for a material with linear dispersion, the Landau levels are given by $E_n=v_D \sqrt{2e \hbar B n}$, where $v_D$ is the Dirac velocity. This provides a less stringent condition for the extreme quantum limit at 10 Oe ($T<4$ K for $v_D=3 \times 10^5$ m/s). Recently, it has been calculated that $\beta$-Ag$_2$Te is indeed a material with a linear dispersion \cite{Zhang2011c}. 

Apart from a single Landau level explaining linear MR, also inhomogeneities in disordered conductors can give rise to linear MR. This is described by the model of Parish and Littlewood \cite{Parish2003,Parish2005}. Regions with higher conductivity create deviations of the direction of the applied current and classical resistor network models show that also in this case a Hall voltage admixture occurs, giving rise to a linear MR \cite{Parish2003,Parish2005}. The Hall signal mixes in such a way that the MR is still symmetric in field. The magnitude of the magnetoresistance from this effect will not exceed the maximum local Hall resistance in the inhomogeneous system. Note, that also the classical inhomogeneity model was used to explain the $\beta$-Ag$_2$Te linear MR \cite{Parish2003}.

The nature of the linear magnetoresistance in topological insulators is not clear yet. In the past few years a lot of research has focused on unraveling the mechanism behind this linear magnetoresistance and to investigate which model applies and whether or not linear MR is a signature of the topological surface states. Furthermore, unravelling the origin of the linear magnetoresistance can have important consequences for spintronics, as the observed magnetoresistance is already large in amplitude and survives even up to room temperature. 

\subsection{Bulk \textit{vs} surface}
Qu $et$ $al.$ \cite{Qu2010} found a magnetoresistance of ~1.7\% at 0.3 K in a magnetic field of 14 T in Bi$_2$Te$_3$ crystals. The observed dependence on perpendicular field is linear, starting already from a very small magnetic field of 30 G. The in-plane MR is also linear, but the effect is a factor of 2 smaller than for perpendicularly applied field. The authors attributed the MR to the coupling of the spin of the bulk electrons to the magnetic field. It is interesting to note that the linear magnetoresistance (LMR) in this case was only observed in the non-metallic samples, while the metallic sample showed quadratic MR. Furthermore, besides the LMR also Shubnikov-de Haas oscillations were observed for the out-of-plane field.

The angular dependence of the LMR was investigated \cite{Tang2011,He2012}. By changing the direction of the field from out-of-plane to in-plane, the LMR decreased significantly and a $|\cos(\theta)|$ relation was found between the LMR and angle of the field. This indicates that the LMR has some 2D character. Tang $et$ $al.$ \cite{Tang2011} deduced the position of the Fermi energy from the 2D sheet carrier density from the SdH oscillations and found this to be consistent with surface states, althought the expected $\pi$ Berry phase could not be clearly resolved. He $et$ $al.$ \cite{He2012} found that the LMR disappears for ultrathin films. This could be related to the gap opening of the surfaces state due to the coupling of the two surfaces with each other \cite{Zhang2010}.

While it is tempting to attribute the LMR to the topological surface states, as bulk transport generally has a 3D character, Cao $et$ $al.$ \cite{Cao2012} have shown in their highly doped Bi$_2$Se$_3$ samples that the observed 2D Shubnikov-de Haas oscillations can originate from the bulk. It is possible that the individual quintuple layers in the bulk all act as 2D parallel electron systems. Solely observing the $|\cos(\theta)|$ is thus not sufficient to attribute the magnetotransport effect to the surface states.

Gao $et$ $al.$ investigated the MR as function of electric gate voltage in Bi$_2$Se$_3$ sheets \cite{Gao2012}. They found that by increasing the negative gate voltage, the quadratic MR changes to a LMR and also the effect of WAL enhances. Comparing this to the temperature dependence of the resistance, a peak appears in the metallic temperature dependence by increasing the gate. This indicates that two components contribute to the charge transport. The thermal excitation of carriers is attributed to an impurity band in the bulk and the metallic component to a state in the gap. It was also assumed that the impurity band gives a quadratic MR, while the gap state has a linear contribution to the MR. Fitting of the MR and the temperature dependence results in consistent values for the gap and bulk state resistance values. By fitting the WAL effect with the HLN equation, Eq. (\ref{eq:HLN}), they found that $\alpha$ increases from -0.6 to -1.2 with increasing gate voltage. Therefore, they attributed the metallic component to the topological surface state. 
However, again it is difficult to distinguish whether the WAL is due the 2D surface states or that the whole sample acts as a 2D system, as is noted in section 2.
\begin{figure}[h!]
\centering
\includegraphics[width=0.5\textwidth]{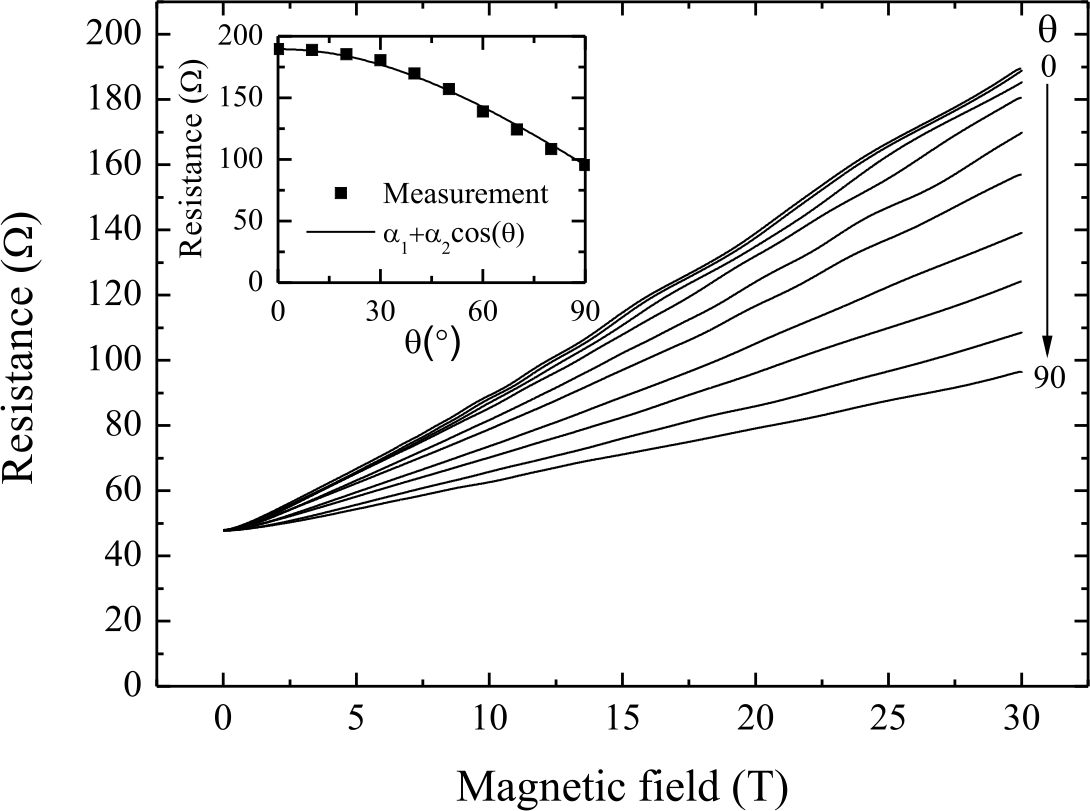}\vspace{-5pt}
\caption{High field measurement of the magnetoresistance of a thin flake of Bi$_{2}$Te$_{3}$. The field was applied from out-of-plane (top curve) to in-plane (bottom curve) in steps of 10 degrees. The inset shows a cosine fit to the data. This dataset was published in Ref. \cite{Veldhorst2012a} after substracting the linear backgrounds.} 
\label{fig:LMR}
\end{figure}

An important observation in the discussion whether 2D effects originate from the bulk or the topological surface is shown in Fig. \ref{fig:LMR}. The dependence of the MR on the applied magnetic field direction suggests 2D behavior (although the MR does not vanish for in plane fields, see also Ref. \cite{Qu2010}). However, from analyzing Hall measurements and Shubnikov de Haas oscillations, it can be shown that the conductance of the bulk of this sample is larger than the conductance from the surface \cite{Veldhorst2012a}. Therefore, when a large 2D effect is then seen on the scale of the total conductace, it seems unlikely that the LMR in this case originates from the surface states alone.

\subsection{Quantum \textit{vs} classical effect}
Attempts have been made \cite{Tang2011,He2012,Wang2012} to explain the observation of LMR by the quantum magnetoresistance model of Abrikosov. The disappearance of the LMR for ultrathin films indicates a relation to the gapless linear dispersion. The low temperature dependence of the observed LMR also points towards Abrikosov's model. However, as Abrikosov's model applies to the extreme quantum limit, it is interesting to note that not all LMR observations \cite{He2012,Wang2012} coincide with observations of Shubnikov-de Haas oscillations. This could indicate that the systems are far away from the extreme quantum limit or that the bulk transport shunts the quantum oscillations. The carrier density in the topological insulator samples are relatively large and for realistic magnetic field values more than one Landau level is usually occupied. If Abrikosov's model is used to explain the LMR in these topological insulators, it is important to know up to which filling factor Abrikosov's model could still apply.

To explain the LMR observed in Bi$_2$Se$_3$ nanosheets \cite{Tang2011}, Wang and Lei \cite{Wang2012a} derive a different model for the LMR using the balance-equation approach \cite{Lei1985}. They found that for a 2D system with linear dispersion and a non-zero g factor a LMR can arise. The model shows that the LMR can occur for fields much lower than $\mu B=1$, it actually requires that the Landau levels overlap. When the mobility decreases, the required field for SdH oscillations increases, whereas the LMR gets even stronger. For higher mobilities, SdH oscillations start to appear and the LMR weakens, because in this regime the overlap of the Landau levels decreases. This also accounts for the low temperature dependence of the LMR, as with higher temperature the Landau levels will broaden, increasing the overlap. It is clear that the system is far away from the extreme quantum limit, in contrast to Abrikosov's model.  By investigating the density dependence of the LMR, the model of Wang and Lei can also be distinguished from Abrikosov's model. The latter predicts an $R_{xx} \propto N^{-s}$ behavior where $s=2$, while the former predicts $s=1$. A positive g factor is required for a positive LMR. The question arises whether Wang and Lei's model could also explain the strength of the LMR, as some reports indicate that the surface g factor could be quite low \cite{Xiong2012a}. This model was derived for Bi$_2$Se$_3$ nanosheets \cite{Tang2011}, but could deviate for other topological insulator samples. It assumes a linear dispersion, but ARPES measurements have shown that Bi$_2$Te$_3$ deviates from a linear dispersion. The bulk contribution was assumed to be negligible, although in many transport measurements the bulk is still prominent. It would be interesting to see whether the model by Wang and Lei \cite{Wang2012a} would still be linear when generalized beyond a linear dispersion and in the presence of bulk shunts.

Alternatively, Assaf $et$ $al.$ \cite{Assaf2012} used a modified HLN model to explain the observed LMR. Their extended HLN model takes the WAL effect and a classical $B^2$ component into account. In a certain magnetic field range, the logarithmic WAL component cancels the $B^2$ component, leaving a linear MR dependence. According to this model, the quadratic component should be distinguishable from a truly linear MR for higher fields, so it would therefore be interesting to investigate whether this idea is still consistent with measurements at even higher fields.

A number of reports \cite{Wang2012,Tang2011,Gao2012,Assaf2012,Wang2012,Zhang2011} mention that the Parish-Littlewood model also could cause a LMR. Although Assaf $et$ $al.$ \cite{Assaf2012} indicate that their samples contain a lot defects, they exclude the possibility that this model explains the LMR, as there is a disagreement between the range of the mobility and the range of the LMR. Others \cite{Wang2012,Wang2012a} also indicate that it is unlikely that the model by Parish and Littlewood explains the observed LMR, as the topological insulator samples have a high quality crystal structure and no structural inhomogeneities are observed. Although Tang $et$ $al.$ \cite{Tang2011} also do not expect structural inhomogeneities, they indicate that electronic inhomogeneities cannot be excluded, which has been confirmed by the observation of potential fluctuations in scanning tunneling measurements \cite{Beidenkopf2011}. More research is therefore required to investigate whether the model by Parish and Littlewood could apply to these topological insulator samples. Expecting a relatively large size of electrodes or contacts in some of the reported transport experiments it would also be useful to study more mundane explanations of LMR such as a non-homogeneous current injection into an otherwise homogenous sample, whereby a transverse Hall component can easily be picked up \cite{Pippard1989}.

\section{Topological insulators with proximity induced superconductivity}
The combination of superconductivity with spin-orbit coupling opens many new exciting research directions, including the realization of a new emergent particle: the Majorana fermion. This particle, which is its own antiparticle, emerges from the electron-hole symmetry of superconducting condensation and the spin-momentum locking in the topological insulator \cite{Fu2008,Fu2009}. The high potential of this particle in quantum computation led to many new proposals in various material systems, such as topological insulator/superconductor structures \cite{Fu2008,Fu2009c,Law2009}, spin-triplet $p$-wave superconductors \cite{Kitaev2001,Read2000,Bolech2007,Sengupta2001,Kraus2009}, helical superconductors \cite{Sato2009c}, topological superconductors \cite{Sasaki2011,Heumen2012,Fu2010c,Hsieh2012b,Yamakage}, semiconductor/metals with strong orbit coupling in combination with superconductors or superconductors with strong spin orbit coupling \cite{Sato2010c,Potter2010,Sau2010,Alicea2010}. In the latter category the most well-known device is inducing superconductivity in nanowires structures with strong Zeeman and Rashba fields \cite{Sau2010,Alicea2010,Lutchyn2010,Oreg2010,Klinovaja2012,Tewari2011,Lutcgyn2011,Golub2011,Flensberg2010,Romito2011,Bena2012,Potter2011a}. First signatures of the Majorana fermion are revealed  \cite{Lutchyn2010,Oreg2010,Mourik2012,Reich2012,Das2012} but in order to test all its peculiar properties the realization of superconducting interference devices will be a prerequisite \cite{Alicea2012,Beenakker2012}.

\subsection{Supercurrents through TI structures}
Shortly after the discovery of the Bi-compound topological insulators, observations of supercurrents in superconductor - topological insulator structures have been reported \cite{Sacepe2011,Veldhorst2012a,Qu2012,Zhang2011,Veldhorst2012b,Williams2012,Zareapour2012}, as well as the coexistence of superconductivity and topological surface states \cite{Wang20122}. Several fabrication techniques have been used such as epitaxial film growth and mechanical exfoliation for the topological insulator structure. Superconducting electrodes are deposited on top resulting in lateral junctions. Sandwich-type junctions have been realized by Qu $et$ $al.$ \cite{Qu2012} using a clever fabrication method based on exfoliated flakes. Using an oxidized Si substrate, Sac\'{e}p\'{e} $et$ $al.$ \cite{Sacepe2011} demonstrated gate tunable supercurrents. The change in the $I_cR_N$ product is of the order of 15\% by applying a gate voltage between -80 V and +50 V. Structures based on exfoliation are therefore flexible and have high potential. High quality topological insulators with small thicknesses and smooth surfaces can be realized that allow the construction of multiple devices on a single flake \cite{Veldhorst2012a}. 
\begin{figure}
\centering
\includegraphics[width=0.4\textwidth]{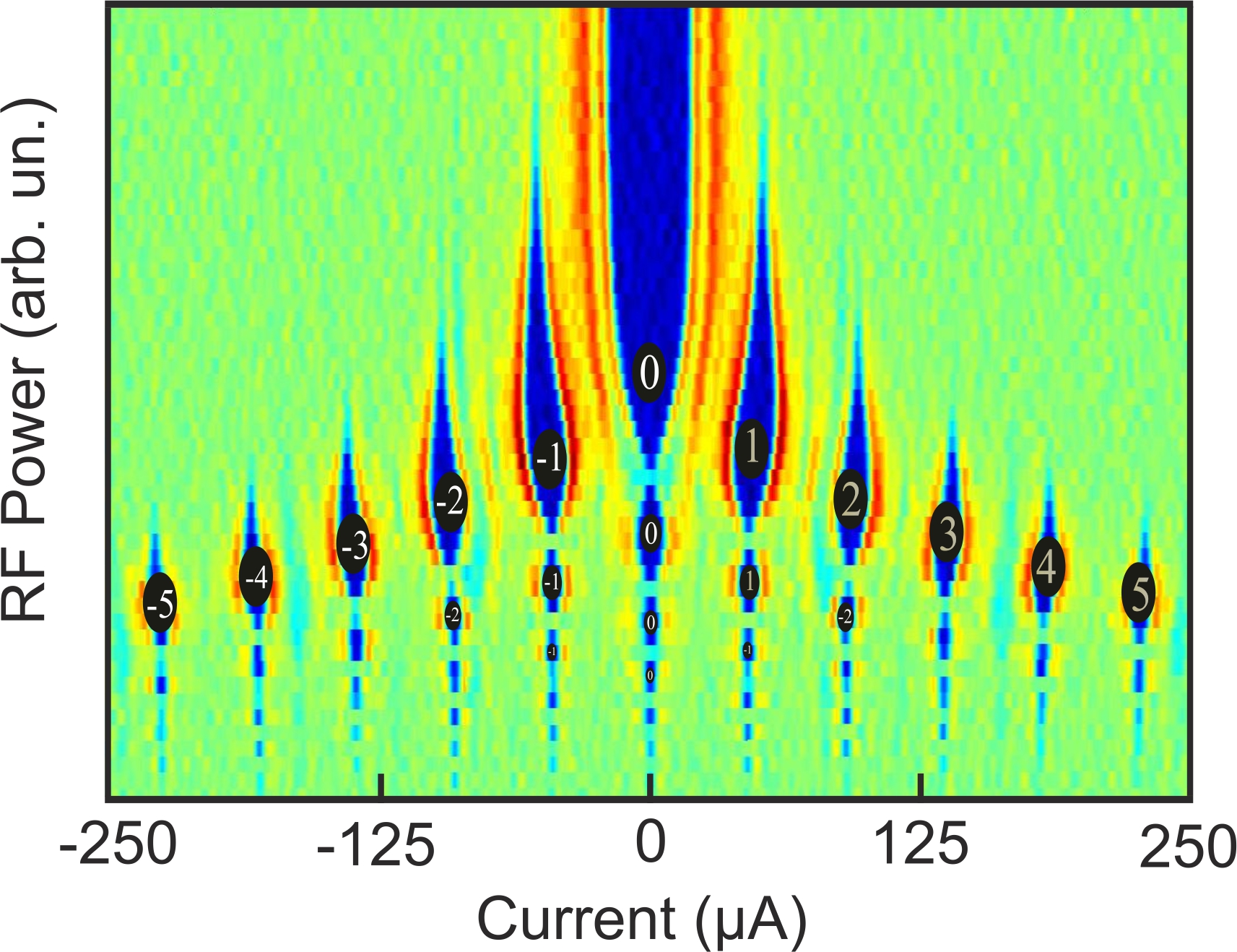}\vspace{-5pt}
\caption{Bessel-peacock color plot of the conductance after irradiation with microwaves (10.0 GHz). From \cite{Veldhorst2012a}.} 
\label{fig:SS}
\end{figure}

The characteristic $I_cR_N$ product observed so far is well below the characteristic voltage scale of $\pi \Delta /e$, e.g. 5 mV for superconducting niobium electrodes. Although mechanisms as scattering and electron-hole decoherence due to junction lengths longer than $\xi$ can lower the $I_cR_N$ product, these junctions often have high interface transparencies and junction lengths comparable to $\xi$. In these systems, an important mechanism lowering the $I_cR_N$ product is the presence of bulk conduction. Normal state transport is dominated by the bulk states which are spread over the entire sample. As the electrodes are fabricated on top, proximity induced superconductivity will be the strongest on the top, and the bulk is effectively shunting the junction $R_N$ significantly, thereby reducing the $I_cR_N$ product. Veldhorst $et$ $al.$ \cite{Veldhorst2012a} obtained a bulk mean free path shorter than the junction length while the system is in the ballistic limit and concluded that the discrimination between bulk and surface states can be so strong that the top topological surface state dominates the superconducting transport, while the bulk states dominate the normal state. 
\begin{figure*}
\centering
\includegraphics[width=0.9\textwidth]{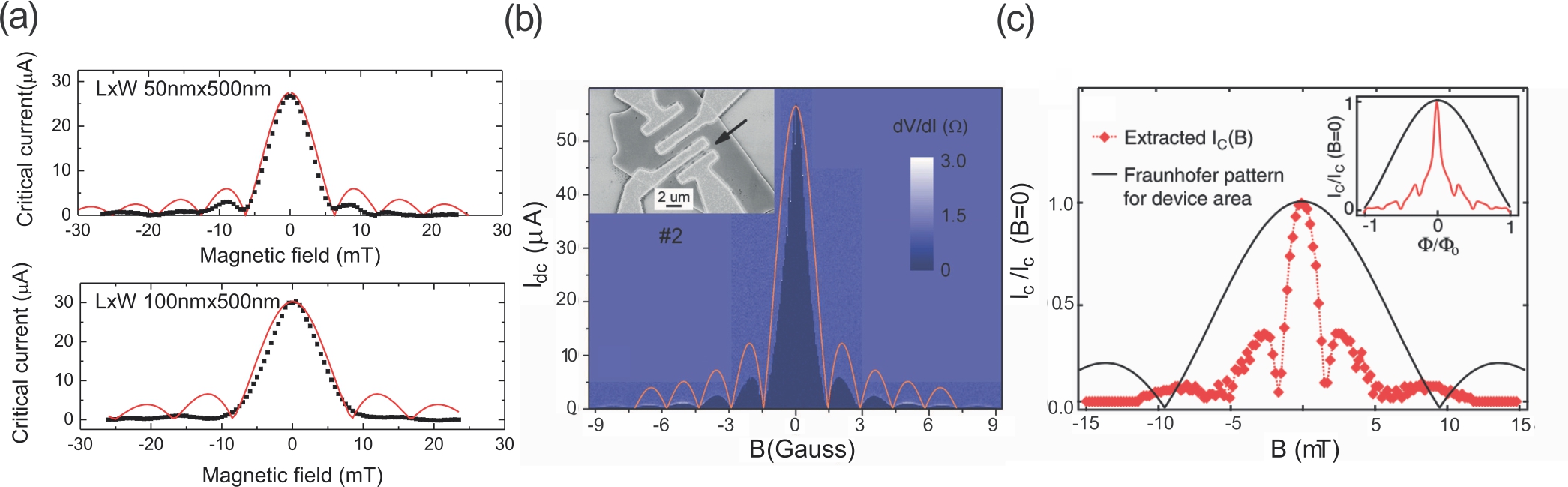}
\caption{Critical current modulation by magnetic field of Josephson junction with topological insulator interlayers. (a) From \cite{Veldhorst2012a}. (b) From \cite{Qu2012}. (c) From \cite{Williams2012}. While the Josephson junctions fabricated by Qu $et$ $al.$ \cite{Qu2012} (b) have standard Fraunhofer patterns, the junctions from Veldhorst $et$ $al.$ \cite{Veldhorst2012a} (a) and Williams $et$ $al.$ \cite{Williams2012} (c) show deviations. Deviations occur both with respect to the field period (which is smaller than expected from the effective junction area) and in the shape.} 
\label{fig:FP}
\end{figure*}

It is an intriguing question why this discrimination is so strong. Besides geometrical effects, strong anisotropy of the band structure of Bi-based topological insulators is a possible origin. The ratio of the superconducting coherence length $\xi_z/\xi_{xy}$  in the topological insulator depends on the effective mass and the potential energy difference along the $xy$-direction and $z$-direction. From the expression of the coherence-length tensor for anisotropic superconductors as described in Ref. \cite{Meuth1988}, we find for the proximity coherence length in the non-superconducting material
\begin{equation}
\frac{\xi_{z}}{\xi_{xy}}\propto \frac{m_{xy}}{m_{z}} \frac{k_{z}}{k_{xy}} \propto \frac{m_{xy}}{m_{z}} \frac{a}{c},
\end{equation}
where the $x$, $y$ and $z$ axis are as defined in Fig. \ref{fig:x}. The superconductivity is induced in the surface states in the $xy-$plane. $a$ and $c$ are the lattice constants of the unit cell along the $x$ (or $y$) and $z$ direction, respectively. Using the valence band effective masses $m_{xy}=0.18$ and $m_{z}=0.84$ for Bi$_{2}$Te$_{3}$ \cite{Yavorsky2011} in conjunction with the lattice constant differences, we find a ratio $\xi_z/\xi_{xy}$ of 0.15. Hence, the band structure itself causes a strong anisotropic proximity effect. An additional argument for strong anisotropy is the observation of the layered quantum Hall effect in Bi$_2$Te$_3$ \cite{Cao2012}. Furthermore, the effective parameter that determines the suppression of $I_cR_N$ in S-N-S junctions (whether ballistic or diffusive, and including barriers at interfaces) is given by $\gamma_{eff}=(\frac{L}{\xi})^2\frac{R_B}{R_N}$, with the boundary resistance $R_B$ and the interlayer resistance $R_N$ \cite{Golubov2004}. From this suppression parameter it can already be seen that a low interlayer resistance with respect to the boundary resistance gives a high $\gamma_{eff}$. The $I_cR_N$ is then suppressed more strongly for the bulk than for the surface channel. Finally, topological effects might render the interfaces to the surface states intrinsically transparent. These intriguing options open a new exciting research direction and give the opportunity to have surface states dominating the supercurrent in topological insulators even in the presence of substantial bulk conductivity.
 
Successive Andreev reflections allow quasiparticles to escape from the superconducting gap in the voltage state. These multiple Andreev reflection (MAR) processes cause structures in the current voltage characteristics at voltages $eV=\alpha\Delta/n$, with $n$ an integer and $\alpha=2$ for standard Cooper pair tunneling and $\alpha=1$ for single electron tunneling mediated by Majorana fermions. Experimentally, Zhang $et$ $al.$ \cite{Zhang2011} have reported evidences for MAR for junctions that are in the standard $\alpha=2$ regime (as expected for 3D junctions). Remarkably, in this experiment only steps are observed for even $n$.

\subsection{Current-phase relationship}
Topological protection causes interfaces to be highly transparent due to Klein tunneling \cite{Fu2008,Fu2009} and absence of backscattering for perpendicular incidence. This protection can change the standard current-phase relationship in Josephson junctions from $\sin({\phi/\alpha})$ with $\alpha=1$ to $\alpha=2$.The systems is then gapless at zero energy, and there exist a Majorana bound state when the phase difference $\phi$ across the junction equals $\pi$. Many theoretical proposals have been put forward to reveal this current-phase relationship. These proposals are based on the $ac$ and $dc$ Josephson effects and superconducting quantum interference devices with topological interlayers.  

The $ac$ Josephson effect has been observed by Veldhorst $et$ $al.$ \cite{Veldhorst2012a}, see Fig. \ref{fig:SS}. These Josephson junctions show clear steps when irradiated with microwaves, demonstrating the Josephson nature of the supercurrent and from the spacing between the steps it can be concluded that these measurements are in the $\alpha=1$ regime. Future strategies to enhance the fraction of $\sin({\phi}/2)$ tunneling involve non-equilibrium measurements \cite{Badiane2011} and the suppression of non-perpendicular trajectories in the junctions \cite{Snelder2012}.   

The $dc$ Josephson effect causes a modulation of the superconducting critical current in an applied magnetic field. In the limit of infinite width and a homogenous current density distribution the magnetic field dependence of the critical current is the Fraunhofer sinc function:
\begin{eqnarray}
I_{c}\left(\Phi_{0} \right) &=& I_{c}\left(0 \right) \dfrac{\sin\left(\pi \Phi/ \alpha \Phi_{0}\right)}{\pi \Phi/\alpha \Phi_{0}}
\label{eq:FP}
\end{eqnarray}
Here, $\Phi_0$ is the superconducting flux quantum. Fig. \ref{fig:FP} shows typical critical current modulation patterns observed so far. The Josephson junctions fabricated by Qu $et$ $al.$ \cite{Qu2012} are well described by Eq. (\ref{eq:FP}) with $\alpha=1$, while Veldhorst $et$ $al.$ \cite{Veldhorst2012a} and Williams $et$ $al.$ \cite{Williams2012} report deviations. 

\begin{figure}
\centering
\includegraphics[width=0.48\textwidth]{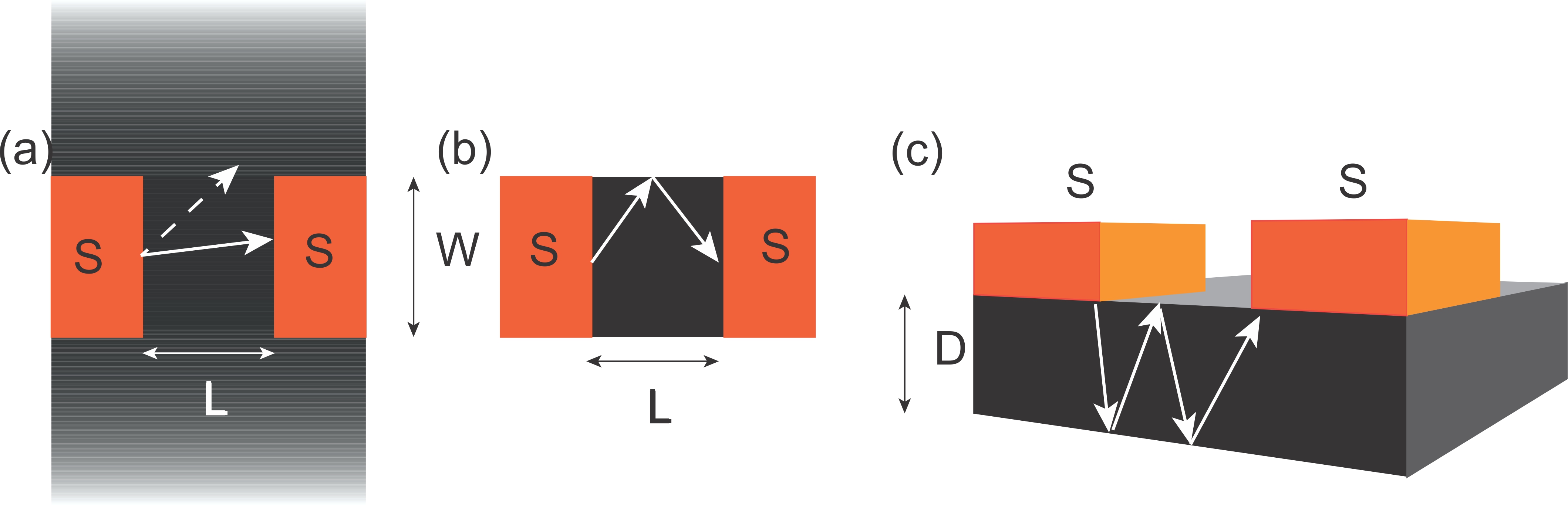}\vspace{-5pt}
\caption{(a) Geometry of S-TI-S junctions as used in the calculation of Ref. \cite{Barzykin1999} to determine the critical current modulation by applied magnetic fields. The dotted arrow shows the path of an electron that never goes to the other superconductor but leaves the junction. In the middle of the superconductor the angle for which the electron still reach the other superconductor is larger than at the edges. (b) Geometry as used in the calculation of Ref \cite{Ledermann1999}. At the boundaries the electrons are specular reflected. (c) The geometry as used in the calculation of Ref. \cite{Mohammadkhani2008}.} 
\label{fig:1}
\end{figure}

Deviations in the field dependence can result from flux focusing effects and geometrical inhomogeneities, e.g. pinholes result in a slower decay of the side lobes, while lower current densities at the edges result in faster decay of the side lobes. Furthermore, having a finite width changes the field dependence, which becomes significant when $W\approx L $, the situation for most experiments so far.  The ratio between the length $L$ and the width $W$ of the junctions can change both the magnitude and period of the diffraction pattern \cite{Heida1998}. The period increases from $\Phi_0$ to $2\Phi_0$ for $L/W \rightarrow \infty$ \cite{Barzykin1999} when the junction edges are `open', as in Fig. \ref{fig:1}(a). In this scenario, the side lobes will decrease more rapidly. 

When specular reflection occurs at the edge of the junction, such as in Fig. \ref{fig:1}(b), the $2\Phi_0$ crossover occurs at smaller aspect ratios\cite{Ledermann1999}: $L/W \sim 1$. The ratio that is important in this scenario is the distance between the Josephson vortex and the range of nonlocal electrodynamics, determined by the thermal length $\xi_{N}$. As long as the range of nonlocal electrodynamics is smaller than the distance between vortices, the $\Phi_{0}$ period remains. For a larger range or strong nonlocality the period becomes $2\Phi_{0}$ due to boundary effects. Including a finite tunneling barrier to this geometry mainly causes a sharpening of the Fraunhofer pattern and the side lobs are flattened \cite{Sheehy2003}. This flattening is also characteristic for SNS junctions in the diffusive limit \cite{Bergeret2008}. A third situation occurs when the magnetic field is parallel to the surface but perpendicular to the current as shown in Fig. \ref{fig:1}(c). This changes the Fraunhofer pattern drastically \cite{Mohammadkhani2008}, including irregular periods, periods smaller than the flux quantum and non-zero minima depending on the $L/D$ ratio. 
The experimentally realized dc SQUIDs by Veldhorst $et$ $al.$ \cite{Veldhorst2012b} based on junctions where the critical current field dependency deviates from the standard Fraunhofer pattern \cite{Veldhorst2012a} show standard fluxoid quantization and suggest that the junction deviations are due to geometrical effects.
 
The observation of Fraunhofer patterns in topological Josephson junctions unambiguously shows the development of the superconducting proximity effect in the topological insulator. The deviations from the standard Fraunhofer form an interesting platform to search for new phenomena in these junctions. However, these junctions are nanosized, are often ballistic, and multiple bands can contribute to the conduction, resulting in complex scenarios. Still, based on these Josephson junctions SQUIDs can be fabricated that can discriminate between geometrical effects and transport mediated by Majorana fermions.

\subsection{dc SQUIDs with unconventional current phase relationships}
\begin{figure}
\centering
\includegraphics[width=0.45\textwidth]{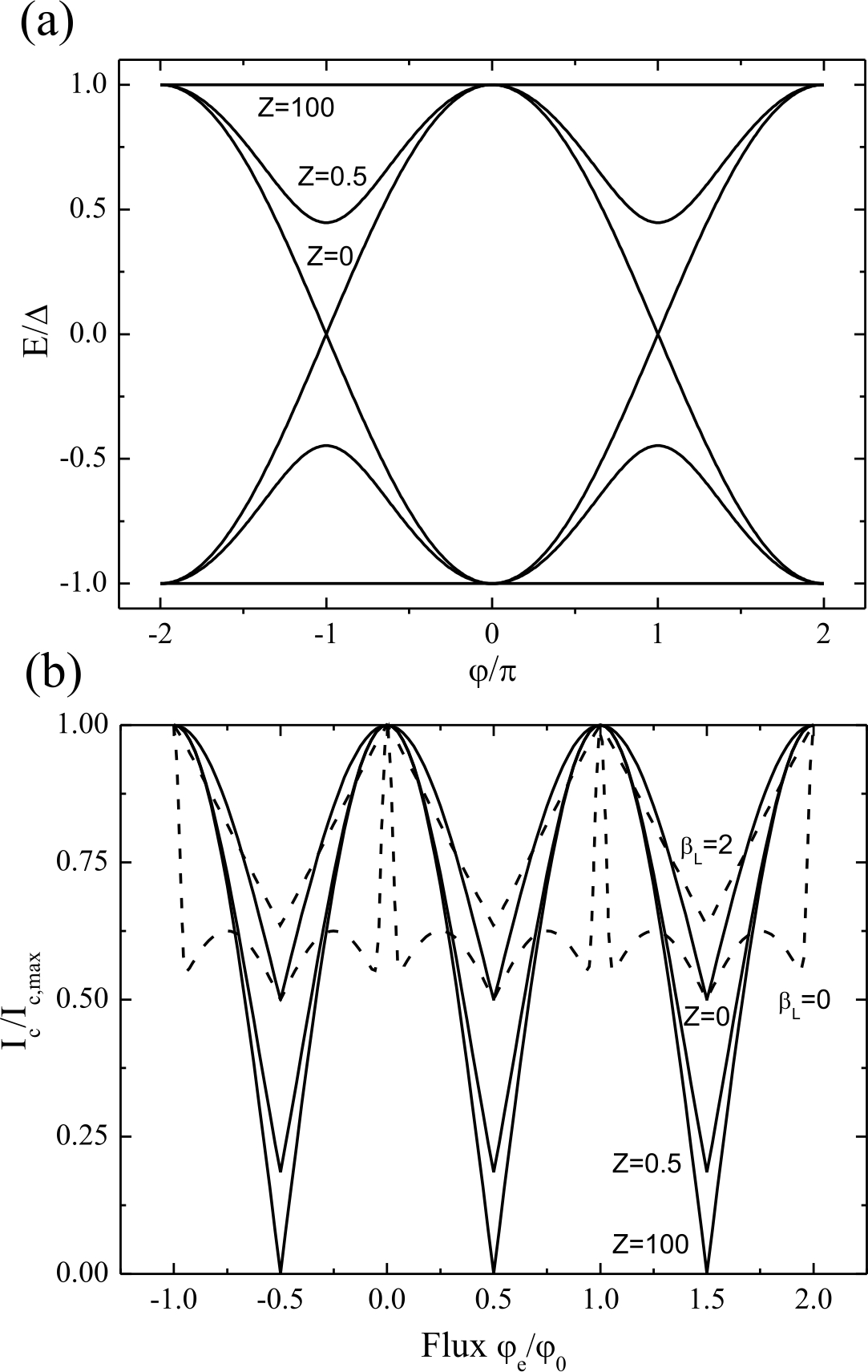}
\caption{SQUIDs including transparant junctions. (a) Current-phase relationship of junctions with different interface transparancies ($Z=0$, 0.5 and 100). (b) Resulting SQUID characteristics. Dashed lines include the peaked current phase relationship as considered in the phenomenological model introduced by Williams $et$ $al.$ \cite{Williams2012}, with the screening parameter $\beta_L=0$ and $2$. These results hold in the presence of relaxation mechanisms and are independent on junction homogeneities and are thereby strong signatures of unconventional current-phase relationships.} 
\label{fig:SQUIDmod1}
\end{figure}
Due to the appearance of Majorana fermions single electron tunneling occurs and the resulting current-phase relation of Josephson junctions can become $4\pi$-periodic. The $4\pi$-periodic current-phase relationship of topological Josephson junctions has its signatures in superconducting loops \cite{Heck2011}. In the absence of relaxation mechanisms, standard fluxoid quantization is doubled. Unfortunately, mechanisms such as quasiparticle poisoning and quantum phase slips will drive the system easily to standard fluxoid quantization. However, as shown by Veldhorst $et$ $al.$ \cite{Veldhorst2012c}, even in this regime the unusual current phase relation of the individual junctions alter the standard SQUID modulation characteristics. This model is based on imposing fluxoid quantization on a superconducting loop interrupted by two Josephson junctions described with the resistivily and capacitively shunted junction (RCSJ) model. To further illustrate this model we show the scenario of full relaxation in Fig. \ref{fig:SQUIDmod1}, where the junctions have been modeled by the Blonder-Tinkham-Klapwijk (BTK) approach \cite{Blonder1982}. The resulting bound states are described by $E=\pm \sqrt{\frac{cos(\phi/2)^2+Z^2}{1+Z^2}}$. Here, $Z$ is the BTK barrier strength of the barriers $I$. These results also represent bound states of superconductor - topological insulator Josephson junctions, where the factor $Z$ effectively describes the scattering resulting from finite angle incidence with momentum mismatches and the appearance of a magnetic gap. In the standard SQUID scenario, the Josephson junctions have low transparency (corresponding to $Z=100$ in Fig. \ref{fig:SQUIDmod1}) and can be described with a $\sin(\phi)$ current-phase relationship. In that case the dc SQUID has a sinusoidal critical current modulation as a function of magnetic field and standard fluxoid quantization $\Phi_0=h/2e$. Lowering the barrier strength, increasing the interface transparency, results in an incomplete critical current modulation \cite{Veldhorst2012c}. This incomplete modulation is a strong signature, since it is independent on the junction homogeneity and survives up to the regime of energy relaxation (where doubled fluxoid quantization is absent \cite{Fu2009}).

We also included the current-phase relationship as imposed phenomenologically by Williams $et$ $al.$ \cite{Williams2012}. This model assumes a current-phase relation that peaks when the relative phase over the junction equals $\pi$. In the regime of small self-induced flux, described by the screening parameter $\beta_L=2\pi L I_c /\Phi_0$, the characteristic spikes are also present in the dc SQUID characteristics. Increasing $\beta_L$ smears out the spikes, resulting in a triangular critical current modulation, see Fig. \ref{fig:SQUIDmod1}. We conclude this section by noting that dc SQUIDs are ideal candidates to measure the unusual current-phase relationships of superconductor-topological insulator structures even in the presence of junction inhomogeneities and relaxation mechanisms. These dc SQUIDs can be used to test whether the critical current field dependences of topological junctions are due to intrinsic or extrinsic effects. 

This work is supported by the Netherlands Organization for Scientific Research (NWO) through VIDI and VICI grants, and by the Dutch Foundation for Fundamental Research on Matter (FOM).


\begin{thebibliography}{[10]}
\providecommand{\WileyBibTextsc}{}
\let\textsc\WileyBibTextsc
\providecommand{\othercit}{}
\providecommand{\jr}[1]{#1}
\providecommand{\etal}{~et~al.}

\bibitem{Fu2007a}
 \textsc{L.~Fu},  \textsc{C. L.~Kane},  and  \textsc{E. J.~Mele},
 \jr{Phys. Rev. Lett} \textbf{98}(10), 106803 (2007).

\bibitem{Fu2007b}
 \textsc{L.~Fu} and  \textsc{C. L.~Kane},
 \jr{Phys. Rev. B} \textbf{76}(4), 045302 (2007).

\bibitem{Zhang2009} H. Zhang \textit{et al.}, Nat. Phys. \textbf{5}, 438 (2009).

\bibitem{Sato2010} T. Sato \textit{et al.}, Phys. Rev. Lett. \textbf{105}, 136802 (2010).
 
\bibitem{Kuroda2010} K. Kuroda \textit{et al.}, Phys. Rev. Lett. \textbf{105}, 146801 (2010).
 
\bibitem{Chen2010b} Y. L. Chen \textit{et al.}, Phys. Rev. Lett. \textbf{105}, 266401 (2010).
 
\bibitem{Souma2012} S. Souma \textit{et al.}, Phys. Rev. Lett. \textbf{108}, 116801 (2012).
 
\bibitem{Kuroda2012} K. Kuroda \textit{et al.}, Phys. Rev. Lett. \textbf{108}, 206803 (2012).
\bibitem{Xu2010} S.-Y. Xu et al., arXiv:1007.5111 (unpublished).
\bibitem{Ren2010} Z. Ren, A. A. Taskin, S. Sasaki, K. Segawa, Y. Ando, Phys. Rev. B \textbf{82}, 241306 (2010). 
\bibitem{Xiong2011} J. Xiong, A. C. Petersen, D. Qu, Y.S. Hor, R. J. Cava, N.P. Ong, Physica E \textbf{44}, 917 (2012).

\bibitem{Ren2011} Z. Ren, A. A. Taskin, S. Sasaki, K. Segawa, Y. Ando, Phys. Rev. B \textbf{84}, 165311 (2011). 

\bibitem{Kitaev2001} A. Y. Kitaev, Phys. -Usp. \textbf{44} 131 (2001).
 
\bibitem{Fu2008} L. Fu, C.L. Kane, Phys. Rev. Lett. \textbf{100}, 096407 (2008).
 
\bibitem{Sasaki2011} S. Sasaki \textit{et al.}, Phys. Rev. Lett. \textbf{107}, 217001 (2011).
 
\bibitem{Heumen2012} E. van Heumen \textit{et al.}, arXiv: 1110.4406 (2012).
\bibitem{Kim2011}
 \textsc{Y.\,S. Kim},  \textsc{M.~Brahlek},  \textsc{N.~Bansal},
  \textsc{E.~Edrey},  \textsc{G.\,A. Kapilevich},  \textsc{K.~Iida},
  \textsc{M.~Tanimura},  \textsc{Y.~Horibe},  \textsc{S.\,W. Cheong},  and
  \textsc{S.~Oh},
 \jr{Phys. Rev. B} \textbf{84}(7), 073109 (2011).
 
\bibitem{Liu2011}
 \textsc{M.~Liu},  \textsc{C.\,Z. Chang},  \textsc{Z.~Zhang},
  \textsc{Y.~Zhang},  \textsc{W.~Ruan},  \textsc{K.~He},  \textsc{L.\,l. Wang},
   \textsc{X.~Chen},  \textsc{J.\,F. Jia},  \textsc{S.\,C. Zhang},
  \textsc{Q.\,K. Xue},  \textsc{X.~Ma},  and  \textsc{Y.~Wang},
 \jr{Phys. Rev. B} \textbf{83}(16), 165440 (2011). 
 
\bibitem{Hikami1980}
 \textsc{S.~Hikami},  \textsc{A.~Larkin},  and  \textsc{Y.~Nagaoka},
 \jr{Prog. Theor. Phys.} \textbf{63}(2), 707--710 (1980).

\bibitem{He2011}
 \textsc{H.\,T. He},  \textsc{G.~Wang},  \textsc{T.~Zhang},  \textsc{I.\,K.
  Sou},  \textsc{G.\,K.\,L. Wong},  \textsc{J.\,N. Wang},  \textsc{H.\,Z. Lu},
  \textsc{S.\,Q. Shen},  and  \textsc{F.\,C. Zhang},
 \jr{Phys. Rev. Lett.} \textbf{106}(16), 166805 (2011).

\bibitem{Zhang2011a}
 \textsc{G.~Zhang},  \textsc{H.~Qin},  \textsc{J.~Chen},  \textsc{X.~He},
  \textsc{L.~Lu},  \textsc{Y.~Li},  and  \textsc{K.~Wu},
 \jr{Adv. Funct. Mater.} \textbf{21}(12), 2351--2355 (2011).

\bibitem{Steinberg2011}
 \textsc{H.~Steinberg},  \textsc{J.B. Lalöe},  \textsc{V.~Fatemi},
  \textsc{J.S. Moodera},  and  \textsc{P.~Jarillo-Herrero},
 \jr{Phys. Rev. B} \textbf{84}(23), 233101 (2011).

\bibitem{Checkelsky2011}
 \textsc{J.\,G. Checkelsky},  \textsc{Y.\,S. Hor},  \textsc{R.\,J. Cava},  and
  \textsc{N.\,P. Ong},
 \jr{Phys. Rev. Lett.} \textbf{106}(19), 196801 (2011).

\bibitem{Cha2012a}
 \textsc{J.\,J. Cha},  \textsc{M.~Claassen},  \textsc{D.~Kong},  \textsc{S.\,S.
  Hong},  \textsc{K.\,J. Koski},  \textsc{X.\,L. Qi},  and
  \textsc{Y.~Cui},
 \jr{Nano Lett.} \textbf{12}(8), 4355--4359 (2012).
 
 \bibitem{Iordanskii1994}
 \textsc{S.~Iordanskii},  \textsc{Y.~Lyanda-Geller},  and
  \textsc{G.~Pikus},
 \jr{JETP Letters} \textbf{60}(3), 206 (1994).

\bibitem{Meijer2005}
 \textsc{F.\,E. Meijer},  \textsc{A.\,F. Morpurgo},  \textsc{T.\,M. Klapwijk},
  and  \textsc{J.~Nitta},
 \jr{Phys. Rev. Lett.} \textbf{94}(18), 186805 (2005).

\bibitem{Wang2011}
 \textsc{J.~Wang},  \textsc{A.\,M. DaSilva},  \textsc{C.\,Z. Chang},
  \textsc{K.~He},  \textsc{J.\,K. Jain},  \textsc{N.~Samarth},  \textsc{X.\,C.
  Ma},  \textsc{Q.\,K. Xue},  and  \textsc{M.\,H.\,W. Chan},
 \jr{Phys. Rev. B} \textbf{83}(24), 245438 (2011).

\bibitem{Takagaki2012}
 \textsc{Y.~Takagaki},  \textsc{B.~Jenichen},  \textsc{U.~Jahn},
  \textsc{M.~Ramsteiner},  and  \textsc{K.\,J. Friedland},
 \jr{Phys. Rev. B} \textbf{85}(11), 115314 (2012).

\bibitem{Zhang2012}
 \textsc{S.~Zhang},  \textsc{L.~Yan},  \textsc{J.~Qi},  \textsc{M.~Zhuo},
  \textsc{Y.\,Q. Wang},  \textsc{R.~Prasankumar},  \textsc{Q.~Jia},  and
  \textsc{S.~Picraux},
 \jr{Thin Solid Films} \textbf{520}(21), 6459--6462 (2012).

\bibitem{Chen2010}
 \textsc{J.~Chen},  \textsc{H.\,J. Qin},  \textsc{F.~Yang},  \textsc{J.~Liu},
  \textsc{T.~Guan},  \textsc{F.\,M. Qu},  \textsc{G.\,H. Zhang},
  \textsc{J.\,R. Shi},  \textsc{X.\,C. Xie},  \textsc{C.\,L. Yang},
  \textsc{K.\,H. Wu},  \textsc{Y.\,Q. Li},  and  \textsc{L.~Lu},
 \jr{Phys. Rev. Lett.} \textbf{105}(17), 176602 (2010).

\bibitem{Chen2011}
 \textsc{J.~Chen},  \textsc{X.\,Y. He},  \textsc{K.\,H. Wu},  \textsc{Z.\,Q.
  Ji},  \textsc{L.~Lu},  \textsc{J.\,R. Shi},  \textsc{J.\,H. Smet},  and
  \textsc{Y.\,Q. Li},
 \jr{Phys. Rev. B} \textbf{83}(24), 241304 (2011).

\bibitem{Taskin2012}
 \textsc{A.\,A. Taskin},  \textsc{S.~Sasaki},  \textsc{K.~Segawa},  and
  \textsc{Y.~Ando},
 \jr{Phys. Rev. Lett.} \textbf{109}(6), 066803 (2012).

\bibitem{Gao2012} B.F. Gao, P.Gehring, M. Burghard, and K. Kern, Appl. Phys. Lett. \textbf{100}, 212402 (2012). 


\bibitem{Zhang2012b}
 \textsc{H.\,B. Zhang},  \textsc{H.\,L. Yu},  \textsc{D.\,H. Bao},
  \textsc{S.\,W. Li},  \textsc{C.\,X. Wang},  and  \textsc{G.\,W. Yang},
 \jr{Phys. Rev. B} \textbf{86}(7), 075102 (2012).

\bibitem{Lu2011}
 \textsc{H.\,Z. Lu} and  \textsc{S.\,Q. Shen},
 \jr{Phys. Rev. B} \textbf{84}(12), 125138 (2011).

\bibitem{Garate2012}
 \textsc{I.~Garate} and  \textsc{L.~Glazman},
 \jr{Phys. Rev. B} \textbf{86}(3), 035422 (2012).

\bibitem{Steinberg2010}
 \textsc{H.~Steinberg},  \textsc{D.\,R. Gardner},  \textsc{Y.\,S. Lee},  and
  \textsc{P.~Jarillo-Herrero},
 \jr{Nano Lett.} \textbf{10}(12), 5032--5036 (2010).

\bibitem{Lu2011a}
 \textsc{H.\,Z. Lu},  \textsc{J.~Shi},  and  \textsc{S.\,Q. Shen},
 \jr{Phys. Rev. Lett.} \textbf{107}(7), 076801 (2011).

\bibitem{Cha2012}
 \textsc{J.\,J. Cha},  \textsc{D.~Kong},  \textsc{S.\,S. Hong},  \textsc{J.\,G.
  Analytis},  \textsc{K.~Lai},  and  \textsc{Y.~Cui},
 \jr{Nano Lett.} \textbf{12}(2), 1107--1111 (2012).

\bibitem{Liu2012}
 \textsc{M.~Liu},  \textsc{J.~Zhang},  \textsc{C.\,Z. Chang},
  \textsc{Z.~Zhang},  \textsc{X.~Feng},  \textsc{K.~Li},  \textsc{K.~He},
  \textsc{L.\,l. Wang},  \textsc{X.~Chen},  \textsc{X.~Dai},  \textsc{Z.~Fang},
   \textsc{Q.\,K. Xue},  \textsc{X.~Ma},  and  \textsc{Y.~Wang},
 \jr{Phys. Rev. Lett.} \textbf{108}(3), 036805 (2012).

\bibitem{Zhang2010}
 \textsc{Y.~Zhang},  \textsc{K.~He},  \textsc{C.\,Z. Chang},  \textsc{C.\,L.
  Song},  \textsc{L.\,L. Wang},  \textsc{X.~Chen},  \textsc{J.\,F. Jia},
  \textsc{Z.~Fang},  \textsc{X.~Dai},  \textsc{W.\,Y. Shan},  \textsc{S.\,Q.
  Shen},  \textsc{Q.~Niu},  \textsc{X.\,L. Qi},  \textsc{S.\,C. Zhang},
  \textsc{X.\,C. Ma},  and  \textsc{Q.\,K. Xue},
 \jr{Nat Phys} \textbf{6}(8), 584--588 (2010).

\bibitem{Liu2009}
 \textsc{Q.~Liu},  \textsc{C.\,X. Liu},  \textsc{C.~Xu},  \textsc{X.\,L. Qi},
  and  \textsc{S.\,C. Zhang},
 \jr{Phys. Rev. Lett.} \textbf{102}(15), 156603 (2009).

\bibitem{Chen2010a}
 \textsc{Y.\,L. Chen},  \textsc{J.\,H. Chu},  \textsc{J.\,G. Analytis},
  \textsc{Z.\,K. Liu},  \textsc{K.~Igarashi},  \textsc{H.\,H. Kuo},
  \textsc{X.\,L. Qi},  \textsc{S.\,K. Mo},  \textsc{R.\,G. Moore},
  \textsc{D.\,H. Lu},  \textsc{M.~Hashimoto},  \textsc{T.~Sasagawa},
  \textsc{S.\,C. Zhang},  \textsc{I.\,R. Fisher},  \textsc{Z.~Hussain},  and
  \textsc{Z.\,X. Shen},
 \jr{Science} \textbf{329}(5992), 659--662 (2010).

\bibitem{Bergmann1984}
 \textsc{G.~Bergmann},
 \jr{Physics Reports} \textbf{107}(1), 1--58 (1984).

\bibitem{Altshuler1979}
 \textsc{B.~Altshuler} and  \textsc{A.~Aronov},
 \jr{Solid State Communications} \textbf{30}(3), 115--117 (1979).

\bibitem{Lee1985}
 \textsc{P.\,A. Lee} and  \textsc{T.\,V. Ramakrishnan},
 \jr{Rev. Mod. Phys.} \textbf{57}(2), 287--337 (1985).

\bibitem{Checkelsky2009}
 \textsc{J.\,G. Checkelsky},  \textsc{Y.\,S. Hor},  \textsc{M.\,H. Liu},
  \textsc{D.\,X. Qu},  \textsc{R.\,J. Cava},  and  \textsc{N.\,P. Ong},
 \jr{Phys. Rev. Lett.} \textbf{103}(24), 246601 (2009).

\bibitem{Beutler1988} 
D.E. Beutler and N. Giordano, Phys. Rev. B \textbf{38}, 8 (1988).

\bibitem{Lin1987} 
J.J. Lin and N. Giordano, Phys Rev. B \textbf{35}, 545 (1987).

\bibitem{Analytis2010a}
 \textsc{J.\,G. Analytis},  \textsc{J.\,H. Chu},  \textsc{Y.~Chen},
  \textsc{F.~Corredor},  \textsc{R.\,D. McDonald},  \textsc{Z.\,X. Shen},  and
  \textsc{I.\,R. Fisher},
 \jr{Phys. Rev. B} \textbf{81}(20), 205407 (2010).

\bibitem{Xiong2012a}
 \textsc{J.~Xiong},  \textsc{Y.~Luo},  \textsc{Y.~Khoo},  \textsc{S.~Jia},
  \textsc{R.\,J. Cava},  and  \textsc{N.\,P. Ong},
 \jr{Phys. Rev. B} \textbf{86}(4), 045314 (2012).

\bibitem{Qu2010}
 D.-X. Qu, Y. S. Hor, J. Xiong, R.J. Cava, and N.P. Ong, Science \textbf{329}, 821 (2010).

\bibitem{Taskin2010}
 A.A. Taskin, K. Segawa, and Y. Ando, Phys. Rev. B \textbf{82}, 121302 (2010).

\bibitem{Analytis2010}
J.G. Analytis, R.D. McDonald, S.C. Riggs, J.-H. Chu, G.S. Boebinger, and I.R. Fisher, Nature Phys. \textbf{10}, 960 (2010).

\bibitem{Sacepe2011}
B. Sac\'{e}p\'{e}, J.B. Oostinga, J. Li, A. Ubaldini, N.J.G. Couto, E. Giannini, and A.F. Morpurgo, Nature Commun. \textbf{2}, 575 (2011).

\bibitem{Brune2011}
 C. Br\"une, C.X. Liu, E.G. Novik, E.M. Hankiewicz, H. Buhmann, Y.L. Chen,X. L.Qi, Z.X. Shen, S. C. Zhang, and L.W. Molenkamp, Phys. Rev. Lett. \textbf{106}, 126803 (2011).

\bibitem{Xiu2011}
F. Xiu, L. He, Y. Wang, L. Cheng, L.-T. Chang, M. Lang, G. Huang, X. Kou, Y. Zhou, X. Jiang, Z. Chen, J. Zou, A. Shailos, and K.L. Wang, Nature Nano. \textbf{6}, 216 (2011).

\bibitem{Taskin2011b}
A.A. Taskin, Z. Ren, S. Sasaki, K. Segawa, Y. Ando, Phys. Rev. Lett. \textbf{107}, 016801 (2011). 

\bibitem{Xiong2012}
J. Xiong, A.C. Petersen, D.-X. Qu, Y.S. Hor, R.J. Cava, and N.P. Ong, Phys. E \textbf{44}, 917 (2012).

\bibitem{Veldhorst2012a}
M. Veldhorst, M. Snelder, M. Hoek, T. Gang, V.K. Guduru, X.L. Wang, U. Zeitler, W.G. van der Wiel, A.A. Golubov, H. Hilgenkamp, and A. Brinkman, Nature Mater. \textbf{11}, 417 (2012).

\bibitem{Cao2012}
 H. Cao, J. Tian, I. Miotkowski, T. Shen, J. Hu, S. Qiao, and Y.P. Chen, Phys. Rev. Lett. \textbf{108}, 216803 (2012).

\bibitem{Ren2011r}
Z. Ren, A.A. Taskin, S. Sasaki, K. Segawa, and Y. Ando, Phys. Rev. B \textbf{84}, 075316 (2011).
\bibitem{Ren2012}
Z. Ren, A.A. Taskin, S. Sasaki, K. Segawa, and Y. Ando, Phys. Rev. B \textbf{85}, 155301 (2012).
\bibitem{Lifshitz1956}
I.M. Lifshitz and A.M. Kosevich, Sov. Phys. JETP \textbf{2}, 636 (1956).
 
\bibitem{Novoselov2005}
K.S. Novoselov, A.K. Geim, S.V. Morozov, D. Jiang, M.I. Katsnelson, I.V. Grigorieva, S.V. Dubonos, A.A. Firsov, Nature \textbf{438}, 197 (2005).

\bibitem{Zhang2005}
Y. Zhang, Y.-W. Tan, H.L. Stormer, P. Kim, Nature \textbf{438}, 201 (2005).

\bibitem{Mikitik2012}
 G.P. Mikitik and Yu.V. Sharlai, Phys. Rev. B \textbf{85}, 033301 (2012).

\bibitem{Taskin2011}
 A.A. Taskin and Y. Ando, Phys. Rev. B \textbf{84}, 035301 (2011). 
 
\bibitem{Wang2012} X. Wang, Y. Du, S. Dou, and C. Zhang, Phys. Rev. Lett. \textbf{108}, 266806 (2012).
 
\bibitem{Tang2011} H. Tang, D. Liang, R.L.J. Qiu, and X.P.A. Gao, ACS Nano \textbf{5}, 7510 (2011).
\bibitem{He2012} H. He, B. Li, H. Liu, X. Guo, Z. Wang, M. Xie, and J. Wang, Appl. Phys. Lett. \textbf{100}, 032105 (2012).

\bibitem{Assaf2012} B.A. Assaf, T. Cardinal, P. Wei, F. Katmis, J.S. Moodera, and D. Heiman, arXiv:1205.4635 (2012). 
\bibitem{Kapitza1928} P. Kapitza, Proc. R. Soc. London, Ser. A \textbf{119}, 358 (1928). 
\bibitem{Abrikosov2000} A.A. Abrikosov, Europhys. Lett. \textbf{49}, 789 (2000). 
\bibitem{Abrikosov1969} A.A. Abrikosov, Sov. Phys. JETP \textbf{29}, 746 (1969).
\bibitem{Abrikosov1998} A.A. Abrikosov, Phys. Rev. B \textbf{58}, 2788 (1998).
\bibitem{Hu2008} J. Hu and T.F. Rosenbaum, Nature Materials \textbf{7}, 697 (2008).
\bibitem{Eto2010} K. Eto, A.A. Taskin, K. Segawa, and Y. Ando, Phys. Rev. B  \textbf{81}, 161202(R) 2010.
\bibitem{Xu1997} R. Xu \textit{et al.}, Nature \textbf{390}, 57 (1997).
\bibitem{Zhang2011c} W. Zhang \textit{et al.}, Phys. Rev. Lett. \textbf{106}, 156808 (2011).
 
\bibitem{Parish2003} M.M. Parish and P.B. Littlewood, Nature \textbf{426}, 162 (2003).
\bibitem{Parish2005} M.M. Parish and P.B. Littlewood, Phys. Rev. B \textbf{72}, 094417 (2005).
\bibitem{Wang2012a} C.M. Wang and X.L. Lei, Phys. Rev. B \textbf{86}, 035442 (2012).
 
\bibitem{Lei1985} X.L. Lei and C.S. Ting, Phys. Rev. B \textbf{30}, 4809 (1984); \textbf{32}, 1112 (1985). 
\bibitem{Beidenkopf2011} H. Beidenkopf \textit{et al.}, Nature Phys. \textbf{7}, 939 (2011).
\bibitem{Pippard1989} A.B. Pippard, Magnetoresistance in Metals, Cambridge Studies in Low Temperature Physics, 1989.
\bibitem{Fu2009} L. Fu, and C.L. Kane, Phys. Rev. B \textbf{79} 161408(R) (2009).
\bibitem{Fu2009c}L. Fu and C. L. Kane, Phys. Rev. Lett. \textbf{102}, 216403 (2009); A. R. Akhmerov, J. Nilsson, C. W. J. Beenakker, Phys. Rev. Lett. \textbf{102}, 216404 (2009); Y. Tanaka, T. Yokoyama, and N. Nagaosa, Phys. Rev. Lett. \textbf{103}, 107002 (2009); J. Linder, Y. Tanaka, T. Yokoyama, A. Sudbo, N. Nagaosa, Phys. Rev. Lett. \textbf{104}, 067001 (2010).
\bibitem{Law2009} K. T. Law, P. A. Lee, and T. K. Ng, Phys. Rev. Lett. \textbf{103}, 237001 (2009).
\bibitem{Read2000}N. Read, D. Green, Phys. Rev. B \textbf{61}, 10267 (2000).
\bibitem{Bolech2007}C. J. Bolech, E. Demler, Phys. Rev. Lett. \textbf{98}, 237002 (2007).
\bibitem{Sengupta2001}K. Sengupta et al., Phys. Rev. B \textbf{63}, 144531 (2001).
\bibitem{Kraus2009} Y. E. Kraus, A. Auerbach, H. A. Fertig, S. H. Simon, Phys. Rev. B \textbf{79}, 134515 (2009).
\bibitem{Sato2009c}M. Sato, S. Fujimoto, Phys. Rev. B \textbf{79}, 094504 (2009); Y. Tanaka, T. Yokoyama, A. V. Balatsky, N. Nagaosa, Phys. Rev. B \textbf{79}, 060505 (2009).
\bibitem{Fu2010c}L. Fu, E. Berg, Phys. Rev. Lett. \textbf{105}, 097001 (2010).
\bibitem{Hsieh2012b} T. H. Hsieh, L. Fu, Phys. Rev. Lett. \textbf{108}, 107005 (2012).
\bibitem{Yamakage}A. Yamakage, K. Yada, M. Sato, Y. Tanaka, Phys. Rev. B \textbf{85}, 180509 (2012).
\bibitem{Sato2010c}M. Sato. Y. Takahashi, S. Fujimoto, Phys. Rev. Lett. \textbf{103}, 020401 (2009); M. Sato,Y. Takahashi, S. Fujimoto, Phys. Rev. B \textbf{82}, 134521 (2010).
\bibitem{Potter2010} A. C. Potter, P.A. Lee, Phys. REv. Lett. \textbf{105}, 227003 (2010); J. Linder, A. Sudbo, Phys. Rev. B \textbf{82}, 08314 (2010); A. Yamakage, Y. Tanaka, N. Nagaosa, Phys. Rev. Lett. \textbf{108}, 087003 (2012).  
\bibitem{Sau2010} J.D. Sau, R.M. Lutchyn, S. Tewari, S. Das Sarma, Phys. Rev. Lett. \textbf{104}, 040502 (2010); T.D. Stanescu, J. D. Sau, R. M. Lutchyn, S. Das. Sarma, Phys. Rev. B \textbf{81}, 241310 (2010).
\bibitem{Alicea2010} J. Alicea, Phys. Rev. B \textbf{81}, 125318 (2010).
\bibitem{Lutchyn2010} R.M. Lutchyn, J. D. Sau, S. Das Sarma, Phys. Rev. Lett. \textbf{105}, 077001 (2010); J. D. Sau, S. Tewari, R. M. Lutchyn, T. D. Stanescu, S. Das Sarma, Phys. Rev. B \textbf{82}, 214509 (2010).
\bibitem{Oreg2010}Y. Oreg, G. Refael, F. von Oppen, Phys. Rev. Lett. \textbf{105}, 214509 (2010).
\bibitem{Klinovaja2012} J. Klinovaja, D. Loss, Phys. Rev. B \textbf{86}, 085408 (2012).
\bibitem{Tewari2011}S. Tewari, T. D. Stanescu, J. D. Sau, S. Das Sarma, New Journal of Physics \textbf{13}, 065004 (2011); J. D. Sau,S. Tewari, S. Das Sarma, arXiv:1111.2054; T. D. Stanescu, R. M. Lutchyn, S. Das Sarma, Phys. Rev. B \textbf{84}, 144522 (2011).
\bibitem{Lutcgyn2011} R. M. Lutchyn, T. D. Stanescu, S. Das Sarma, Phys. Rev. Lett. \textbf{106}, 127001 (2011); R. M. Lutchyn, M. P. A. Fisher, Phys. Rev. B \textbf{84}, 214528 (2011).
\bibitem{Golub2011} A. Golub, I. Kuzmenko, and Y. Avishai, Phys. Rev. Lett. \textbf{107}, 176802 (2011).
\bibitem{Flensberg2010}K. Flensberg, Phys. Rev. B \textbf{82}, 180516 (2010).
\bibitem{Romito2011} A. Romito, J. Alicea, G. Refael,F. von Oppen, arXiv:1110.6193 (2011); J. Alicea, arXiv:1202.1293.
\bibitem{Bena2012}C. Bena, D. Sticlet, and P. Simon, Phys. Rev. Lett. \textbf{108}, 096802 (2012).
\bibitem{Potter2011a} A. C. Potter and P. A. Lee, Phys. Rev. B \textbf{83}, 094525 (2011); B. Zhou, S.-Q. Shen, Phys. Rev. B \textbf{84}, 054532 (2011); K. T. Law and P. A. Lee, Phys. Rev. B \textbf{84},081304 (2011).
\bibitem{Mourik2012} V. Mourik, K. Zuo, S.M. Frolov, S.R. Plissard, E.P.A.M. Bakkers, and L.P. Kouwenhoven, Science \textbf{25}, 336 (2012).
\bibitem{Reich2012}E. S. Reich, Nature News \textbf{483}, 132 (2012); F. Wilczek, Nature \textbf{486}, 195 (2012).
\bibitem{Das2012}A. Das, Y. Ronen, Y. Most, Y. Oreg, M. Heiblum, H. Shtrikman, arXiv:1205.7073; L. P.
Rokhinson, X. Liu, J. K. Furdyna, arXiv:1204.4212; M. T. Deng, C. L. Yu, G. Y. Huang, M. Larsson, P. Caroff, H. Q. Xu, arXiv:1204.4130.
\bibitem{Alicea2012} J. Alicea, Rep. Prog. Phys. \textbf{75}, 076501 (2012).
\bibitem{Beenakker2012} C.W.J. Beenakker, arXiv: 1112.1950v2.
\bibitem{Qu2012} F. Qu, F. Yang, J. Shen, Y. Ding, J. Chen, Z. Ji, G. Liu, J. Fan, X. Jing, C. Yang, L. Lu, Sci Rep., 2: 339 (2012).
\bibitem{Zhang2011} D. Zhang, J. Wang, A. M. DaSilva, J. S. Lee, H. R. Gutierrez, M. H. W. Chan, J. Jain, N. Samarth, PRB \textbf{84}, 165120 (2011).
\bibitem{Veldhorst2012b} M. Veldhorst, C. G. Molenaar, X. L. Wang, H. Hilgenkamp, A. Brinkman, Appl. Phys. Lett. \textbf{100}, 072602 (2012)
\bibitem{Williams2012} J. R. Williams, A. J. Bestwick, P. Gallagher, S. S. Hong, Y. Cui, A. S. Bleich, A. S. Bleich, J. G. Analytis, I. R. Fisher, D. Goldhaber-Gordon, PRL \textbf{109}, 056803 (2012).
\bibitem{Zareapour2012}P. Zareapour et al., Nature Communications \textbf{3},1056 (2012).
\bibitem{Wang20122} M.X. Wang, C. Liu, J.P. Xu, F. Yang, L. Miao, M.Y. Yao, C.L. Gao, C. Shen, X. Ma, X. chen, Z.A. Xu, Y. Liu, S.C. Zhang, D. Qian, J.F. Jia, and Q.K. Xue, Science \textbf{336}, 52 (2012).
\bibitem{Meuth1988} H. Meuth, Phys. Rev. B \textbf{38} (1988).
\bibitem{Yavorsky2011} B. Y. Yavorsky, N. F. Hinsche, I. Mertig, P. Zahn, Phys. Rev. B \textbf{84}, 165208 (2011).
\bibitem{Golubov2004} A.A. Golubov, M.Yu. Kupriyanov, and E. IL'ichev, Rev. Mod Phys. \textbf{76}, 411 (2004).
\bibitem{Badiane2011} D.M. Badiane, M. Houzet, and J.S. Meyer, Phys. Rev. Lett. \textbf{107}, 177002 (2011).
\bibitem{Snelder2012} M. Snelder, M. Veldhorst, A.A. Golubov, A. Brinkman, submitted for publication (2012).
\bibitem{Heida1998} J. P. Heida, B. J. van Wees, T. M. Klapwijk, G. Borghs, Phys. Rev. B \textbf{57}, 9995 (1998).
\bibitem{Barzykin1999} V. Barzykin, A. M. Zagoskin, Superlat. Microstruct. \textbf{25}, 797â€"807 (1999).
\bibitem{Ledermann1999} U. Ledermann, A. L. Fauch\`{e}re, G. Blatter, Phys. Rev. B \textbf{59}, R9027 (1999).
\bibitem{Sheehy2003} D. E. Sheehy, A. M. Zagoskin, Phys. Rev. B \textbf{68}, 144514 (2003). 
\bibitem{Bergeret2008} F.S. Bergeret, J.C. Cuevas, J Low Temp Phys \textbf{153}, 304 (2008).
\bibitem{Mohammadkhani2008} G. Mohammadkhani, M. Zareyan, Y. M. Blanter, Phys. Rev. B \textbf{77}, 014520 (2008). 
\bibitem{Heck2011} B. van Heck, F. Hassler, A.R. Akhmerov, and C.W.J. Beenakker, Phys. Rev. B \textbf{84}, 180502(R) (2011).
\bibitem{Veldhorst2012c} M. Veldhorst, C. G. Molenaar, C. J. M. Verwijs, H. Hilgenkamp, A. Brinkman, Phys. Rev. B \textbf{86}, 024509 (2012).
\bibitem{Blonder1982} G.E. Blonder, M. Tinkham, T.M. Klapwijk, Phys. Rev. B \textbf{25}, 4515 (1982).

\end{thebibliography}
\end{document}